\documentclass[structabstract]{aa}  
\usepackage{graphicx}
\usepackage{txfonts}
\usepackage{natbib}
\bibpunct{(}{)}{;}{a}{}{,}
\begin{document}
   \title{Deep \textit{XMM-Newton} observation of the Eta Chamaleontis 
   cluster\thanks{This publication makes use of data products from the 
   Two Micron All Sky Survey, which is a joint project of the University 
   of Massachusetts and the Infrared Processing and Analysis Center/California 
   Institute of Technology, funded by the National Aeronautics and Space 
   Administration and the National Science Foundation.}}

   \subtitle{}

   \author{J. L\'opez-Santiago
          \inst{1}
          \and
          J. F. Albacete Colombo\inst{2}
          \and
          M. A. L\'opez-Garc\'ia\inst{1}
          }

   \institute{Departamento de Astrof\'{\i}sica y Ciencias de la 
              Atm\'osfera, Universidad Complutense de Madrid,
              E-28040 Madrid, Spain\\
              \email{jls@astrax.fis.ucm.es; mal@astrax.fis.ucm.es}
         \and
             Centro Universitario Regional Zona Atl\'antica (CURZA) 
             Universidad Nacional del COMAHUE, Monse\~nor Esandi 
             y Ayacucho (8500), Viedma (Rio Negro), Argentina.\\
             \email{donfaca@gmail.com}
             }

   \date{Received ...; accepted ...}

% \abstract{}{}{}{}{} 
% 5 {} token are mandatory
 
  \abstract
  % context heading (optional)
  % {} leave it empty if necessary  
   {The members of the $\eta$ Chamaleontis cluster are in an evolutionary stage in which 
    disks are rapidly evolving. It also presents some peculiarities, such as the large fraction 
    of binaries and accretion disks, probably related with the cluster formation process. 
    Its proximity makes this stellar group an ideal target 
    for studying the relation between X-ray emission and those stellar parameters. }
  % aims heading (mandatory)
   {The main objective of this work is to determine general X-ray properties of the 
   cluster members in terms of coronal temperature, column density, emission measure, 
   X-ray luminosity and variability. We also aim to establish the relation between the X-ray 
   luminosity of these stars and other stellar parameters, such as effective temperature, 
   binarity, and presence of accretion disks. Finally, a study of flare energies for each 
   flare event detected during the observations and their relation with some stellar 
   parameters is also performed.}
  % methods heading (mandatory)
   {We used proprietary data from a deep \textit{XMM-Newton} EPIC observation pointed at the 
   core of the $\eta$ Chamaleontis cluster. Specific software for the reduction of \textit{XMM-Newton}
   data was used for the analysis of our observation. For the 
   detection of sources in the composed EPIC pn+mos image, we used the wavelet-based code 
   \textit{PWDetect}. General coronal properties were derived from plasma model fitting. 
   X-ray light curves in the 0.3--8.0~keV energy range were generated for each star.}
  % results heading (mandatory)
   {We determined coronal properties and variability of the $\eta$ Chamaleontis members in the 
   \textit{XMM-Newton} EPIC field-of-view. A total of six flare-like events were clearly detected in five 
   different stars. For them, we derived coronal properties during the flare events and pseudo-quiescent
   state separately. In our observations, stars that underwent a flare event have higher X-ray luminosities
   in the pseudo-quiescent state than cluster members with similar spectral type with no indications
   of flaring, independently whether they have an accretion disk or not. Observed flare energies are 
   typical of both pre-main- and main-sequence M stars. We detected no difference between flare energies
   of stars with and without an accretion disk.}
  % conclusions heading (optional), leave it empty if necessary 
   {}


   \maketitle
%
%________________________________________________________________

\section{Introduction}
\label{intro}

The $\eta$~Chamaleontis cluster (hereafter $\eta$~Chamaleontis), named after the 
eponymous B8V star $\eta$ Cha, 
was discovered by \citet{mam99} using \textit{ROSAT} data. The authors proposed 
13 members, twelve of them being X-ray sources. Later, they investigated main 
properties of the sources \citep{mam00} and established a connection between the 
$\eta$~Chamaleontis cluster and the $\epsilon$~Chamaleontis stellar association 
\citep{fri98}. %(discovered by Frink et al. 1998, A\&A, 338, 442). 
They are part of a number of nearly coeval stellar groups that 
includes the TW~Hya association and that probably had their origin in the molecular
complex Scorpius-Centaurus \citep[see][]{mam00}. Its proximity and low stellar
density converts the $\eta$~Chamaleontis in an unique scenario in the solar vicinity.
%Together with the TW~Hya association, these young stellar groups presumably  
%had their origin in the Scorpius-Centaurus (Sco-Cen) giant molecular cloud. 
%The large H\,I filament and dust lane located near $\eta$~Chamaleontis has been 
%identified in the past as part of a superbubble formed by Sco-Cen OB winds and 
%supernova remnants \citep[see][]{mam00}. The passage of the superbubble 
%may have terminated star formation in the $\eta$~Chamaleontis cluster and 
%dispersed its natal molecular gas, converting the cluster in an 
%unique scenario in the solar vicinity. 

%$\eta$~Chamaleontis presents some particularities. 
With an age of 6--8 Myr, $\eta$~Chamaleontis presents some particularities. 
Its members are in an evolutionary stage in which disks are rapidly evolving
(see Table~\ref{tab0}) %probably due to processes related to planet building and 
and can provide further constraints on inner disk lifetimes \citep{hai05}.
It has been found that several cluster members 
are in fast transition from classical T-Tauri stars (CTTS-like) to debris disks 
\citep{meg05,sic09}. Besides, the binary fraction observed in the cluster is twice 
higher than that found for field stars and dense clusters \citep{lyo04}, but similar 
to the binary fraction in Taurus. This result may suggest that there is a relation 
between the binary fraction and the stellar density of star-forming regions.
Some studies of the Initial Mass Function (IMF) of $\eta$~Chamaleontis 
\citep[e.g.][]{mor07} reveal that the cluster mass function %is consistent with the 
%IMF of rich young open clusters in the highest mass range, but that it 
shows a deficit of low mass stars and brown dwarfs. Nevertheless, \citet{luh04} 
%assures that the absence of brown dwarfs in this cluster
%is statistically consistent with the mass functions measured in star-forming
%regions, that exhibits only 1 or 2 brown dwarfs in stellar samples of the size
%of $\eta$~Chamaleontis. 
indicates that other low-mass star-forming regions as Taurus have 
very few or none brown dwarfs in stellar samples of the size
of $\eta$~Chamaleontis.
However, the deficit of low mass stars in the cluster
is still not well-known. \citet{mor07}, using numerical simulations, 
concluded that the IMF is typical of clusters formed from
a compact configuration and that dynamical interactions can result
in the lost of the original cluster members.
For the age of $\eta$~Chamaleontis, those
stars ejected in the early phases of the cluster formation might reach 
distances of 6 to 10 pc from the cluster center. Thus, one should look for 
these ejected members at angular distances up to 5 degrees from the 
core of the cluster. This scenario has been recently validated by 
\citet{mur10}, who identified four new probable cluster members 
and three possible members.  
With this new four stars, the census of known cluster members grows up to 
twenty two \citep[see][for a recent review]{tor08}. 

The proximity \citep[$d \approx 97$ pc;][]{mam99} and properties of 
$\eta$ Chamaleontis converts this cluster in 
an ideal target for pointed X-ray observations. In this paper, we present a
detailed study of the X-ray properties of cluster members based on a new
deep \textit{XMM-Newton} observation. Details of this observation are 
presented in Section~\ref{obs}, together with the data reduction plan. 
In Section~\ref{corona}, we derive general coronal properties and treat to relate 
them with other stellar properties as binarity and the presence of accretion 
disks. X-ray variability (including flare-like variability) is studied in Section~\ref{variability}.
In that section, we also investigate the relation between flare energies and 
different stellar parameters. Section~\ref{candidates} is dedicated to other 
possible members in the field-of-view of our observation. Notes on particular
sources are given in Section~\ref{notes}. Finally, in Section~\ref{summary}
we briefly summarize the main results of this work.

\section{X-ray observation and data reduction}
\label{obs}

\begin{figure}
   \centering
   \includegraphics[width=9cm]{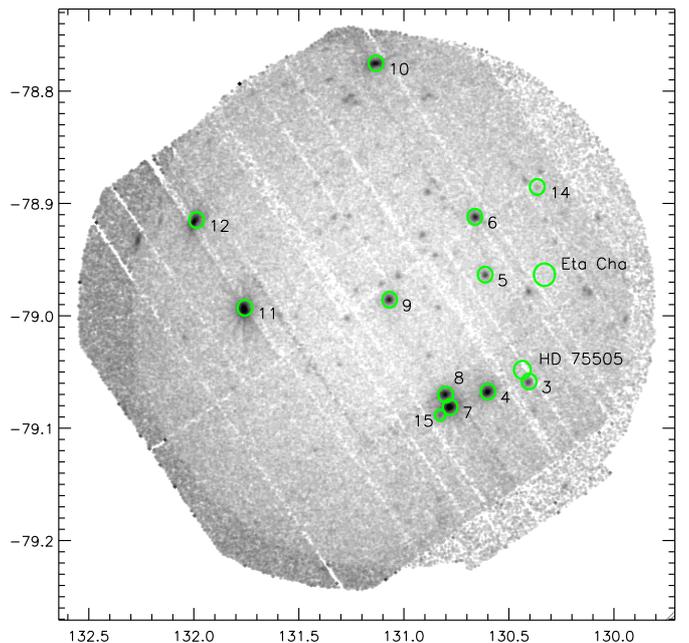}
      \caption{EPIC-pn and mos mosaic of the $\eta$ Chamaleontis
      cluster. Known members of the cluster are marked with a circle and
      identified with its RECX number. The two members not detected in X-rays
      ($\eta$ Cha and HD 75505) are identified with their most common name 
      instead of their RECX identification (RECX 2 and RECX 13, respectively).}
       \label{f1}
   \end{figure}

Our \textit{XMM-Newton} observation of the $\eta$ Chamaleontis cluster (id. 0605950101) 
was performed on a single exposure of 48.3 ks on June 2009. The EPIC was used in Full 
Frame mode with the Thick filter to reduce the contamination of the X-ray signal by visible 
and UV radiation. The effective exposure time of the observation was 46.4 ks in the EPIC-pn
and 48.0 ks in the EPIC-mos. 
We performed a standard reduction using the version 9 of the specific \textit{XMM-Newton} 
reduction package SAS. 
%to derive a table of calibrated events.  Then, different filters were applied to eliminate bad events and noise. 
The observation was not affected by background flaring events and therefore, no time was lost by 
high X-ray variable background. %The final good-time intervals (GTI) events file was used for our study. 
For our study, we used a good-time-intervals events file, i.e cleaned for bad events and pixels and 
noise. 
%
%We created images in the different energy bands for each 
%detector individually. Then, exposure maps were constructed using the EEXMAP routine in the 
%SAS. These maps provide the spatial efficiency of the instruments. By dividing the images by the 
%exposure maps, we corrected for the quantum efficiency, filter transmission, and mirror vignetting. 
%The resultant images were divided by the corresponding effective area to account for the difference 
%in efficiency of the EPIC-pn and EPIC-mos detectors. Finally, the task EMOSAIC was used to 
%construct a combined pn+mos image. The resultant image is shown in Fig.~\ref{f1}. 
The image shown in Fig.~\ref{f1} was created by constructing a combined EPIC-pn+mos 
image using the task EMOSAIC of the SAS. EPIC images were first individually corrected for the 
quantum efficiency, filter transmission and mirror vignetting by dividing by the exposure maps 
obtained with the routine EEXMAP.

%_____________________________________________________________
\begin{table}[!t]
\caption{Spectral type, binarity and disk properties of 
$\eta$ Chamaleontis members in the XMM-Newton field of view.} 
\label{tab0}      
\scriptsize
\centering          
\begin{tabular}{lcccccccc}     % 8 columns 
\hline\hline       
Source & RECX &  Sp.T.\tablefootmark{*} & Binarity\tablefootmark{**}  & 
  Disk type\tablefootmark{\dag} & $\log L_\mathrm{bol}$\tablefootmark{\ddag} \\ 
              &             &             &                &                   & (erg\,s$^{-1}$) \\
\hline                
$\eta$ Cha & 2 & B8 & ? & debris & 35.5 \\ %33.5 \\   
EH Cha & 3   & M3 &   & TO & 32.7 \\ %31.6 \\
EI Cha  &  4   & M1 &   & TO & 32.9 \\ %31.9 \\
EK Cha & 5   & M4 &   & TO & 32.6 \\ % 31.4 \\
EL Cha & 6   & M3  &   &      & 32.7 \\ %31.7 \\
EM Cha & 7  & K6 & K6.9+M1.0 &  & 33.4 \\ %32.4 \\
RS Cha & 8   & A7 + A8 & eclipsing & & 34.9 \\ %33.4 \\
EN Cha & 9   & M4 & M4.4 + M4.7 & TO & 32.6 \\ %31.6 \\
EO Cha & 10 & M0 & & & 33.2 \\ %31.9 \\
EP Cha & 11 & K6 & & class II & 33.4 \\ % 32.3 \\
EQ Cha & 12 & M3 & M3.2+M3.2: & & 32.7 \\ % 32.0 \\
HD 75505 & 13 & A1 & & & 35.1 \\ % 32.9 \\
ES Cha & 14 & M5 & & TO & 32.1 \\ % 31.0 \\
ET Cha & 15 & M3 & & class II & 32.7 \\ % 31.5 \\
\hline                  
\end{tabular}
\tablefoot{
\tablefoottext{*} \citet{tor08}. 
\tablefoottext{**} \citet{sic09}, determined from \citet{lyo04}.
\tablefoottext{\dag} \citet{sic09}. TO are transitional disk objects.
%\tablefoottext{\ddag} Luminosities determined from the 2MASS J magnitude
%and assuming a distance of 97 pc for all the stars.
\tablefoottext{\ddag} Luminosities determined from the spectral type, assuming 
an age of 6 Myr and the \citet{sie00}'s pre-main-sequence model.
}
\end{table}
%_____________________________________________________________

A modified version of the PWDetect code \citep{dam97} was used for the detection of sources in the 
0.3--7.5 keV energy band. We chose the detection threshold $SNR = 5$, which corresponds to 
one possible spurious detection in each field, as obtained from our simulations of the background 
in \textit{XMM-Newton}\footnote{We acknowledge Dr. I. Pillitteri for his help with the \textit{XMM-Newton} 
background simulations}. 
With this threshold, a total of 86 sources were detected in the combined EPIC-pn and mos 
image. A direct visual inspection allowed us to discard 7 multiple detections (sources detected 
more than once by the detection algorithm) plus two spurious detections in the detector borders. 
The complete list of source detections is given in Table~\ref{tabA1}. For each source, we give its position, 
significance of the detection, observed count-rate and flux. This flux was determined by converting
observed fluxes in photons\,cm$^{-2}$\,s$^{-1}$ to erg\,cm$^{-2}$\,s$^{-1}$ using the conversion 
factor $CF = 1.5 \pm 0.2 \times 10^{-9}$ erg\,ph$^{-1}$. This factor was determined  from the 
spectral fitting of the known $\eta$ Chamaleontis members (see Section~\ref{corona}). Note that 
these fluxes are merely indicative. Most of the detected sources are Active Galactic Nuclei 
instead of stellar coronae. To check this fact, we suggest the reader to compare the
fluxes determined for cluster members using the conversion factor (Table~\ref{tabA1})
with the fluxes obtained from the spectral fitting (Table~\ref{tab1}).

The value used by us as detection threshold corresponds to a source count-rate limit of 
completeness $CR = 5.5  \pm 1.5 \times 10^{-4}$ s$^{-1}$ in the EPIC energy band. Assuming 
a typical corona with $kT \sim 0.5-1$ keV, this count-rate corresponds to a flux 
$f_\mathrm{X} \sim 1-3 \times 10^{-15}$ erg\,cm$^{-2}$\,s$^{-1}$, or $L_\mathrm{X} \sim 2 \times 
10^{27}$ erg\,s$^{-1}$ at a distance of 100 pc (an upper limit to the distance of cluster members). 
This assures us that all the cluster members emitting in X-rays have been detected. 
Compared to observations of other clusters and star-forming regions, our observation is one 
order of magnitude deeper in X-ray luminosities. % (see Table~\ref{tab0}).

In Fig.~\ref{f1} we mark with a circle the position of the known cluster members in the
field of view of our observation. 
%Note that, out of the eighteen known members of $\eta$ Chamaleontis \citep[see][]{tor08},
%fourteen fall inside the EPIC field of view in our observation. However, 
Note that two of them were not 
detected: the B star $\eta$~Cha and the A star HD 75505 (see Section~\ref{notes} for details).

\section{General X-ray properties of $\eta$ Chamaleontis members}

\subsection{General properties}
\label{corona}

The fourteen known cluster members in the EPIC field of view include three early-type 
stars: $\eta$ Cha (B8\,V), HD 75505 (A0\,V), and RS Cha (A7\,V) and 11 Li-rich, 
H$_\alpha$ emission-line late-type (K6--M5) stars. The first X--ray characterization 
of this region was performed by \citet{mam99} based on pointed \textit{ROSAT}-HRI 
[0.1-2.4 keV] broadband observations (see Section~\ref{intro} for details). Unfortunately, 
those observations did not allow the authors to perform a complete spectral characterization 
of the detected sources, due to the limited spectral response of  the HRI \citep[see][]{mam00}. 
With the \textit{XMM-Newton} [0.3-10 keV] observation, we are able to give parameters 
of the hot plasma by fitting plasma models to the X-ray spectrum of each star. 

It is important to note here that, for our study, we assume that all the X-ray emission comes from 
the (hot) coronal plasma. This is basically correct for non-accreting stars. However, it has been 
demonstrated that accretion shocks generate dense, relatively hot  plasma that emits in very 
soft X-rays \citep[e.g.][]{sac08,bri10}. Its contribution to the overall stellar X-ray emission is 
negligible in medium and hard X-rays ($kT \ge 0.5$ keV). But, when the accretion rate
is high (of the order of $\sim 10^{-10}$ M$_\odot$\,yr$^{-1}$), the X-ray luminosity from the shocked plasma 
is of the same order of magnitude than the typical X-ray luminosity of a T Tauri star \citep[see][]{sac08}. 
Of the four accreting stars in our sample, only one (RECX 15) shows a high accretion rate \citep[e.g.][]{sic09}.
Nevertheless, it is highly absorbed in soft X-rays (see Table~\ref{tab1} and discussion below).
Thus, the contribution of X-rays produced by the accretion to its spectrum is presumably 
very low, although high resolution observations are needed to investigate this fact.

The EPIC spectra --\,integrated for the whole exposure\,-- were analyzed using a 2T-temperature 
model. In some cases, a third temperature was added to obtain a robust fit at hard (i.e. above 2 keV) 
X-ray energies. We note here that the stars for 
which a 3-T model was used underwent flare-like events during the exposure 
(see Section~\ref{variability}). For them, we performed a more detailed study separating 
the quiescent from the flare state. The results of this study are shown in Table~\ref{tab3} 
and in Section~\ref{variability}.
For the fit, we used the Astrophysical Plasma Emission Code \citep[APEC,][]{smi01}
which is included in the  XSPEC spectral fitting package \citep{arn96, arn04}. APEC is a 
routine that generates spectral models for hot (optically thin) plasmas using the atomic data 
contained in the Astrophysical Plasma Emission Database \citep[APED,][]{2001ASPC..247..161S}. 
We added a multiplicative interstellar absorption model in XSPEC --\,in particular the 
one described in \citet{mor83}\,-- to account for possible absorption due to interstellar and/or
circumstellar material \citep[for a more detailed description of the method we refer the reader 
to][]{lop08}.
%For the fit, we used the XSPEC spectral fitting package 
%\citep{arn96, arn04} in the EPIC-pn, mos1, and mos2 detectors simultaneously. We adopted 
%the Astrophysical Plasma Emission Code \citep[APEC,][]{smi01} included in the XSPEC 
%software. APEC calculates spectral models for hot, optically thin plasmas using 
%the Astrophysical Plasma Emission Database \citep[APED,][]{2001ASPC..247..161S}, 
%that contains the relevant atomic data for calculating both the continuum and line emission.
%Interstellar absorption was taken into account using the interstellar photo-electric absorption 
%cross-sections of \citet{mor83}, also available in XSPEC.

Best-fit parameters for the chosen models were found by $\chi^2$ minimization and are 
shown in Fig.~\ref{fA1}. A word of caution is necessary here: the resulting 
fits as given below are merely formal and indicative. A 2-T model fitting to an observed coronal
spectrum does not mean that we have to deal with two different plasmas at two different 
temperatures. In general, the coronal plasma exhibits a temperature gradient that cannot
be studied using low-resolution X-ray spectra \citep[although see][]{rob05}. Nevertheless, 
it has been shown that the dual-temperature nature of the fits may represent an intrinsic 
property of the coronal spectra of moderately active late-K and M stars \citep{sch90,bri03,lop07}. 
In this scenario, the third (higher) temperature required to fit the hard tail of the X-ray spectrum 
is though to be present only during flare-like events. In most cases, a successful 1-T model 
fitting is a mere reflection of a lack of information due to the low statistics (i.e. count-rate) of 
many sources.

\begin{table*}
\caption{Spectral parameters of known members}
\label{tab1}
\scriptsize
\label{spectra}\centering
\begin{tabular}{l c c c c c c c c c c c c c c}
\hline
RECX & $N_{\rm H}$ & $kT_{\rm 1}$  & $kT_{\rm 2}$ & $kT_{\rm 3}$ & 
$EM_1/EM_2$  & $EM_1/EM_3$ & $Z$ & $\chi^2$ (d.o.f.)  & Unabsorbed $f_\mathrm{X}$  & $\log L_\mathrm{X}$ & $\log L_\mathrm{X}$\tablefootmark{**} \\ 
           & ($\times 10^{21}$ cm$^{-2}$) & (keV) &  (keV) & (keV) & 
 & & ($Z_\odot$)  & & ($\times 10^{-13}$ erg cm$^{-2}$ s$^{-1}$) & (erg s$^{-1}$) & (erg s$^{-1}$) \\
\hline
\noalign{\smallskip}
3  & 0.00$^{+0.31}_{-0.00}$  & 0.63$^{+0.04}_{-0.04}$  & ... & ...& ... & ... & 0.08$^{+0.02}_{-0.02}$ & 1.12 (88) & 0.65$^{+0.06}_{-0.10}$ & 28.9 & 29.2 \\
\noalign{\smallskip}
4  & 0.44$^{+0.12}_{-0.11}$  & 0.28$^{+0.02}_{-0.01}$  & 0.96$^{+0.04}_{-0.03}$ & ... & 0.92 & ... & 0.12$^{+0.02}_{-0.01}$ & 1.29 (402) & 6.61$^{0.71}_{2.97}$ & 29.9 & 30.2 \\
\noalign{\smallskip}
5  & 0.00$^{+0.25}_{-0.00}$  & 0.27$^{+0.07}_{-0.06}$  & 0.74$^{+0.18}_{-0.09}$ & ... & 1.30 & ... & 0.10$^{+0.06}_{-0.03}$ & 0.96 (162) & 0.56$^{+0.04}_{-0.12}$ & 28.8 & 29.1\\
\noalign{\smallskip}
6  & 0.23$^{+0.22}_{-0.19}$  & 0.33$^{+0.04}_{-0.03}$  & 0.95$^{+0.05}_{-0.06}$ & ... & 0.88 & ... & 0.13$^{+0.04}_{-0.03}$ & 0.93 (347) & 2.47$^{+0.35}_{-0.20}$ & 29.4 & 29.6 \\
\noalign{\smallskip}
7  & 0.03$^{+0.13}_{-0.03}$  & 0.33$^{+0.01}_{-0.01}$  & 0.94$^{+0.05}_{-0.04}$ & 2.08$^{+1.42}_{-0.36}$ & 0.75 & 2.21 & 0.19$^{+0.03}_{-0.04}$ & 1.03 (491) & 10.50$^{+0.78}_{-0.50}$ & 30.1 & 30.4 \\
\noalign{\smallskip}
8 & 0.00$^{+0.10}_{-0.00}$  & 0.26$^{+0.06}_{-0.03}$  & 0.76$^{+0.08}_{-0.06}$ & 1.54$^{+0.14}_{-0.13}$ & 0.87 & 0.61 & 0.26$^{+0.05}_{-0.05}$ & 1.28 (353) & 4.05$^{+0.39}_{-0.31}$ & 29.7 & 29.9 \\
\noalign{\smallskip}
9 & 2.89$^{+0.24}_{-0.29}$  & 0.77$^{+0.02}_{-0.04}$  & 2.22$^{+0.18}_{-0.18}$ & ... & 0.47 & ... & 0.30$^{+0.13}_{-0.06}$ & 1.12 (518) & 2.95$^{+0.21}_{-0.38}$ & 29.5 & 28.5 \\
\noalign{\smallskip}
10 & 0.04$^{+0.19}_{-0.04}$  & 0.40$^{+0.09}_{-0.04}$  & 0.98$^{+0.05}_{-0.06}$ & ... & 0.72 & ... & 0.15$^{+0.04}_{-0.04}$ & 0.92 (300) & 4.07$^{+0.39}_{-0.31}$ & 29.7 & 30.0 \\
\noalign{\smallskip}
11 & 0.06$^{+0.01}_{-0.01}$ & 0.25$^{+0.02}_{-0.01}$ & 0.95$^{+0.04}_{-0.04}$ & 2.52$^{+0.45}_{-0.27}$ & 0.90 & 1.20 & 0.21$^{+0.10}_{-0.05}$  & 1.20 (805) & 27.93$^{+1.37}_{-2.83}$ & 30.5 & 30.2 \\
\noalign{\smallskip}
12 & 0.11$^{+0.20}_{-0.11}$ & 0.28$^{+0.02}_{-0.02}$ & 1.00$^{+0.03}_{-0.03}$ & 2.90$^{+3.51}_{-0.99}$ & 0.73 & 2.45 & 0.20$^{+0.10}_{-0.09}$  & ... & 10.90$^{+1.10}_{-0.58}$ & 30.1 & 30.2 \\
\noalign{\smallskip}
14\tablefootmark{*} & ... & ... & ... & ... & ... & ... & ... & ... & ... & ...  & ... \\
\noalign{\smallskip}
15 & 1.27$^{+0.47}_{-0.52}$  & 0.80$^{+0.18}_{-0.09}$  & ... & ...& ... & ... & 0.04$^{+0.02}_{-0.02}$ & 1.05 (64) & 0.31$^{+0.07}_{-0.24}$ & 28.8 & ... \\
\noalign{\smallskip}
%4  & $<0.001$                       &0.27$^{0.01}_{0.01}$  & 0.82$^{0.02}_{0.02}$ &1.25$^{0.09}_{0.09}$ & 0.13$^{0.06}_{0.06}$ & $1.1 \times 10^{-3}$ & 6.32$^{0.07}_{0.07}$ & & 1.3 (400) \\
%\noalign{\smallskip}
%5 & 0.09$\pm$0.03 &0.36$\pm$0.03&...		& ...	 & $<$0.02 &4.5$\times$10$^{-4}$&1.02$\pm$0.03 & & 1.07 (158)\\    
%6 & 0.04$\pm$0.01 &0.31$\pm$0.02&0.84$\pm$0.03	& ...		& 2.2$\pm$0.2	&5.0$\times$10$^{-5}$&3.11$\pm$0.02 & & 0.95 (343)\\
%7 & $<$0.001	   &0.329$\pm$0.008&0.93$\pm$0.02	&2.2$\pm$0.4	& 2.2$\pm$0.2	&1.46$\times$10$^{-3}$&10.8$\pm$0.09 & & 1.04 (480)\\
%8 & 0.13$\pm$0.02 &0.28$\pm$0.02 &0.97$\pm$0.04	&2.2$\pm$0.6	& 0.54$\pm$0.15&4.92$\times$10$^{-3}$&8.82$\pm$0.05 & & 1.3 (351)\\
%9 & 0.42$\pm$0.02 &0.72$\pm$0.02 &2.13$\pm$0.09	&... & 0.40$\pm$0.07&5.1$\times$10$^{-4}$&6.28$\pm$0.05 & & 1.4 (530)\\
%10& 0.14$\pm$0.02 &0.31$\pm$0.02 &0.96$\pm$0.04	&...		& 0.27$\pm$0.09&3.3$\times$10$^{-3}$&8.50$\pm$0.05 & & 1.1 (286) \\
%12& $<$0.0001	   &0.28$\pm$0.01 &0.98$\pm$0.02	&2.5$\pm$0.5	& 0.20$\pm$0.04&1.4$\times$10$^{-3}$&10.8$\pm$0.05 & & 1.1 (222) \\
\hline
\end{tabular}
\tablefoot{
\tablefoottext{*}{The background subtracted spectrum of RECX 14 has only 67 counts. We failed any 
attempt of fitting a plasma model.}
\tablefoottext{**}{X-ray luminosities determined by \citet{mam00} in the 0.1--2.0 \textit{ROSAT} energy band.}
}
\end{table*}

In Table\,\ref{tab1} we give details on the best-fit parameters for each star. As a goodness-of-fit
test, we give the reduced $\chi^2$ and degrees of freedom (d.o.f.) used for the fit. Unabsorbed 
fluxes in the energy band 0.3--8.0 keV are also given. Finally, the X-ray luminosities listed in 
the table were determined using the same distance for all the stars ($d = 97$ pc). It is noticeable
that our luminosities are 0.2--0.3 dex smaller than those determined by \citet{mam00} in a 
smaller energy range, except for EN Cha (RECX 9) and EP Cha (RECX 11) that underwent 
long duration flares during our exposure (see Section~\ref{variability} and Fig.~\ref{f3}).
Nevertheless, it must be said that the authors could not perform model fitting to their
data because of the poor spectral response of the \textit{ROSAT}-HRI. \citet{mam00} then
used a conversion factor to derive X-ray fluxes from the observed count-rates. Besides, 
they could not correct the observed fluxes for absorption. Taking all these factors into 
account, their estimations of the X-ray luminosities of the $\eta$ Chamaleontis members 
were quite reasonable. 

The results presented in Table~\ref{tab1} about the coronal temperature should be interpreted 
with caution, since different effects are contributing to the X-ray spectra. On average, the corona 
of the stars showing no flares during our observation are parametrized by a 2T-model with 
temperatures k$T_1 \approx 0.30$ keV and k$T_2 \approx 0.95$ keV. For stars that 
underwent a flare during the exposure, a third (hotter) component needs to be added to the 
model. For the two stars with lower count-rates (i.e. lower statistics), in which a 1T-model 
was used, the temperature obtained from the fit represents a mean value between  k$T_1$ and k$T_2$, 
weighted by the emission measure. We refer the reader to \citet{lop07} and \citet{cab10} for a more detailed 
discussion on this issue. These results suggest that the X-ray spectrum of these stars during 
their quiescent state can be parametrized by a plasma model typical of both main- and 
pre-main-sequence stars. 
Only the star EN Cha (RECX 9) deviates from this trend (k$T_1 \approx 0.77$ keV, 
k$T_2 \approx 2.22$ keV). The reason is likely that its X-ray spectrum is highly absorbed 
(at longer wavelengths) due to the presence of a circumstellar disk \citep{sic09}, 
rather than it is caused by high X-ray emission produced during a flare-like event.
Note that the star underwent a flare before the beginning of the exposure while
we observed only the decay phase (see Fig.~\ref{f3}), were the plasma is rapidly 
cooling.
The spectral analysis of the known cluster members gave low mean 
column densities, with mean $N_\mathrm{H} = 4.6 \times 10^{20}$ cm$^{-2}$, 
with typical deviation $\sigma = 8.5 \times 10^{20}$ cm$^{-2}$. The observed dispersion 
is due to the high $N_\mathrm{H}$ values found for the sources RECX 9 and 15. Both 
are well-known CTTS Ê\citep[e.g.][]{sic09}. 
%and the disk precense is probable the most plausible explanation of a such high N$_{\rm H}$ values.

\begin{figure*}
\centering
\includegraphics[width=9cm]{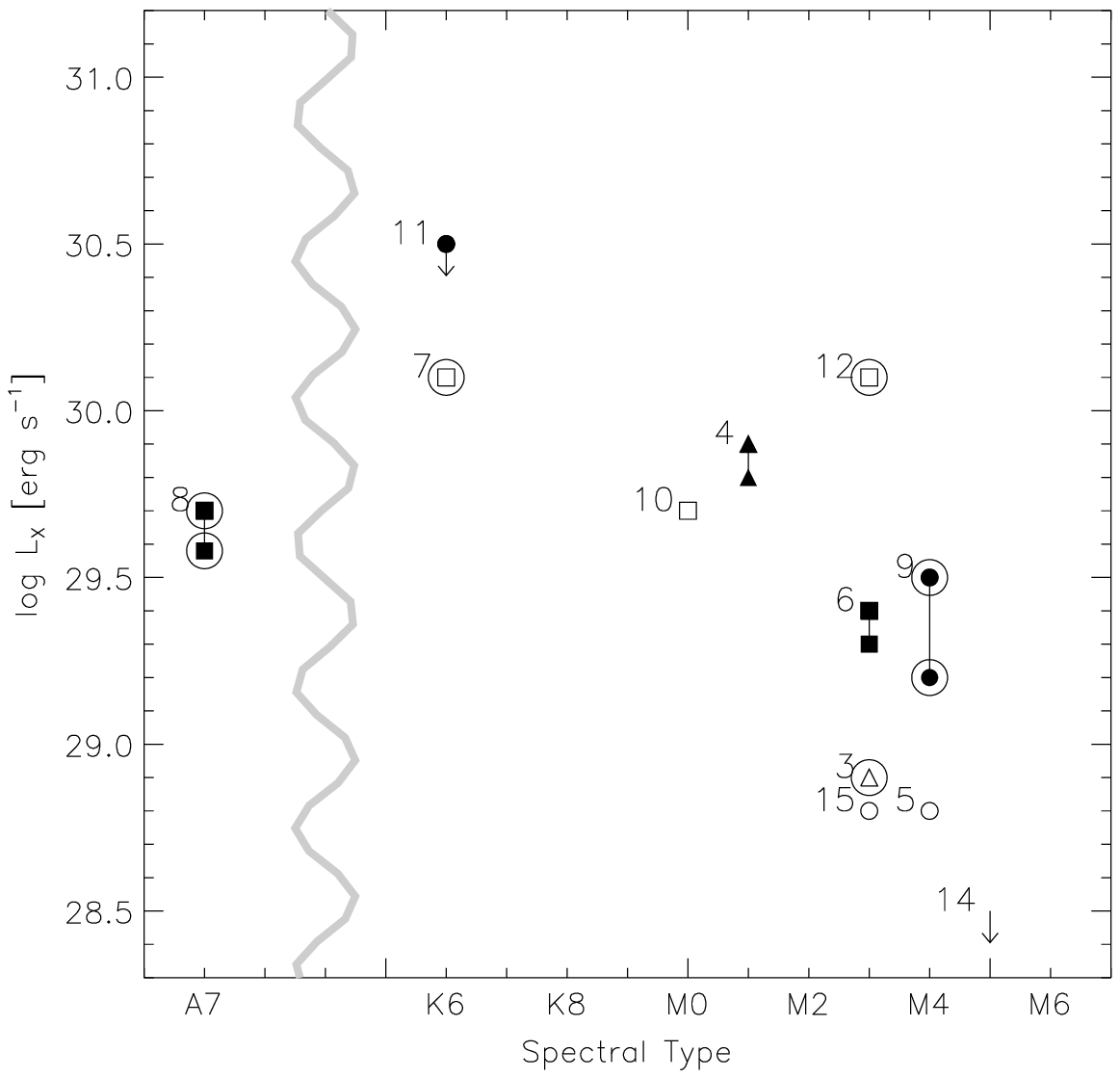}
\includegraphics[width=9cm]{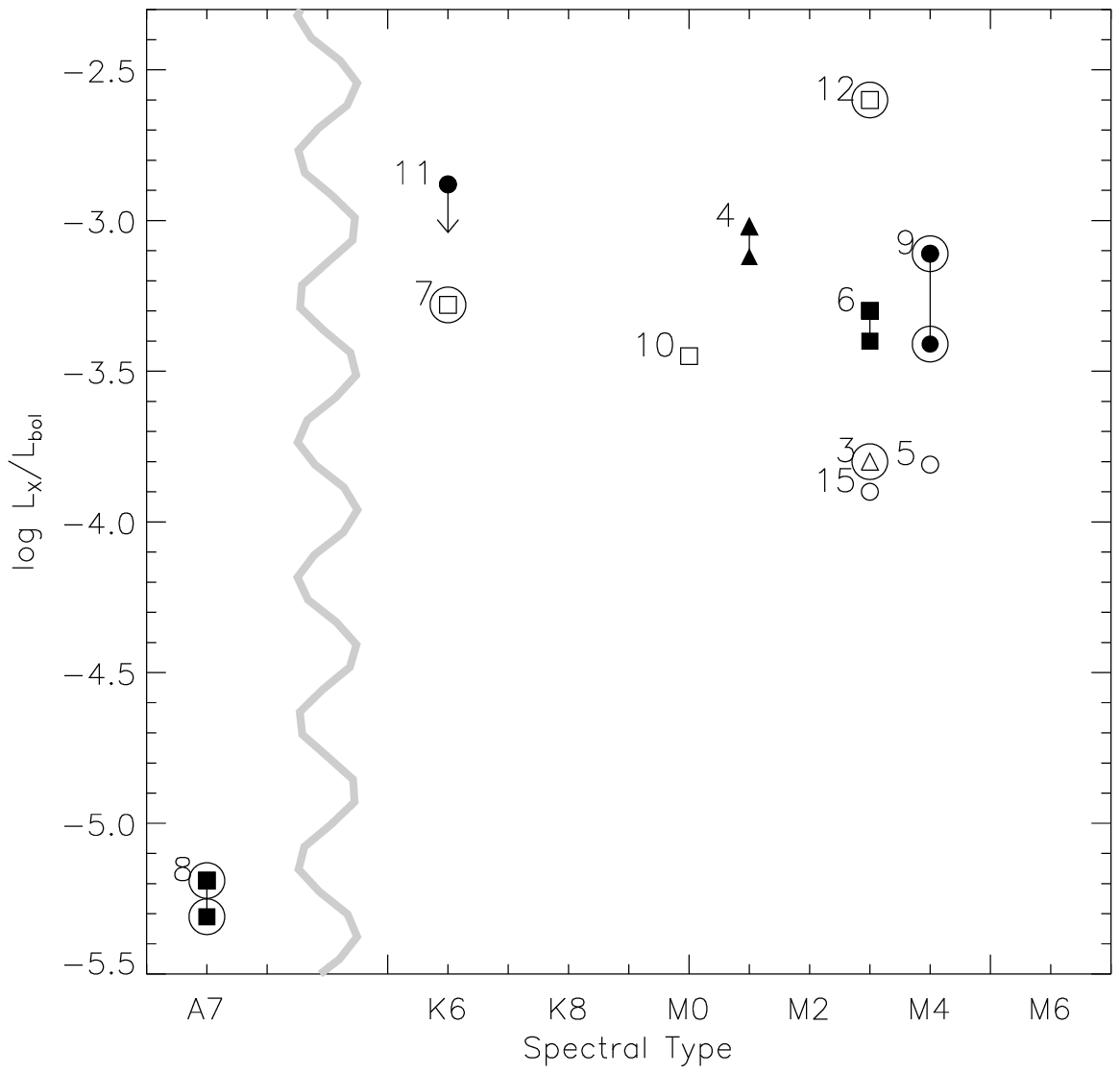}
   \caption{X-ray luminosities of known members of $\eta$ Chamaleontis
                  in the energy band 0.3--8.0 keV. Circles are known classical T Tauri
                  stars, triangles are transitional objects and squares are stars with no 
                  signatures of accretion disks. Filled symbols mark stars that 
                  underwent a flare during our observation. Binary systems are marked
                  with a large circle. Note that RECX 12 was observed, probably, in flare 
                  state during the exposure but was not clearly detected in its light curve 
                  due to its low count-rate (see Section~\ref{corona}).
              }
   \label{f2}
\end{figure*}

To investigate possible correlations between average X-ray luminosity and any stellar 
parameter, such as binarity, mass, or the presence of accretion disks, we used data from 
the literature (see Table~\ref{tab0}).
In Fig~\ref{f2} we plot different symbols for each object type \citep[as classified by][]{sic09}: 
classical T Tauris (CTTSs, circles), transitional disk objects (TO, triangles), and weak-line 
T Tauris (WTTSs, squares). Filled symbols are stars with flare events during our observations. 
For these stars, we plot also the X-ray luminosity in the pseudo-quiescent state (see Section~\ref{flares}). 
Large circles mark binary stars. The upper-limit symbol used for RECX 11 denotes 
that its quiescent-state X-ray emission is probably lower than that determined from our 
observations. Note that, for our study, we assumed the emission in the time-period $20 < t < 30$ ks
to be representative of the quiescent state of this star but this period is between two 
flare events (see Fig.~\ref{f3}).
No particular differences between the X-ray luminosity of binary and single stars or
stars with and without an accretion disk is detected, although the sample is not large 
enough to extract any robust conclusion from this result. The typical trend of 
decreasing X-ray luminosity with decreasing mass \citep{pre05} is observed.
Nevertheless, the stars that underwent a flare during our observations 
have higher luminosities than stars with no flares, even when the contribution of the 
flare is subtracted. Note that EQ Cha (RECX 12) very probably underwent a flare during the 
observations. A 3T-model with k$T_3 = 2.90$ keV was necessary for the fit 
(see Table~\ref{tab1}). This difference in the X-ray luminosities for stars with similar spectral 
types may be explained by an enhancement of the X-ray emission (quiescent) level before the 
flares occur. This explanation should be interpreted with caution, since the statistic in 
terms of number of stars is very poor.

%_____________________________________________________________
%_____________________________________________________________

\subsection{X-ray variability}
\label{variability}

X-ray variability in time scale of days was reported by \citet{mam00} for seven cluster members. 
The authors attributed this variability to flare-like events superposed on moderately variable 
emission, but the poor signal of those observations prevented them to perform a 
quantitative analysis.  In general, moderate variations in scale of a few days are probably 
owed to rotational modulation \citep{mar03}. Examples of it are shown in \citet{cab09} for 
the $\sigma$ Orionis cluster and \citet{fla05} for Orion. Flare-like events are usually short in time
(from a few minutes to a few hours) and imply a large release of energy. In young stars and 
particularly in T Tauri stars, they have been observed in a large variety of shapes 
\citep[e.g.][]{fav05,fra07,lop10}. The \textit{Chandra} and, specially, the \textit{XMM-Newton} 
missions due to its large collector area, are powerful to detect flaring events in stars. In this sense, 
our observation is suitable to detect short-time ($< 30$ minutes) and medium-time (several hours) 
variations.

We extracted light curves from the EPIC-pn for the cluster members revealed in our observation, 
except for RECX 12, for which we used EPIC-mos since it was located in a gap between chips 
in the EPIC-pn. We chose a specific extraction radius for each source, depending on the position 
of the source at the detector, to avoid the lost of counts due to the degradation of the PSF towards 
the borders of the EPIC. A background region was chosen close to the source, in the same 
chip but far from other sources, and then subtracted from the source light curve.  A different 
time binning was also used for each source depending on the source count-rate. The resultant
background-subtracted light curves are shown in Fig.~\ref{f3}. In Table~\ref{tab2}, 
we give details on the parameters for the extraction of the light curves (extraction radius and time 
binning), the type of variability detected and the count-rate increase factor during the flare
in those stars in which a flare was clearly detected. We classified the type of variability as 
\textit{flare}, when flare-like events where observed, or simply \textit{variable}, for other types 
of variability. 
RECX 15 (ET Cha) is revealed above the background only in a small fraction of the 
exposure, around 30 ks from the beginning of the observation. We suspect this 
star underwent a flare an that its quiescent emission is below the detection limit 
of this observation. 

%_____________________________________________________________
%                                             One column Table 
%_____________________________________________________________
%
\begin{table}[!t]
\caption{Characteristics of the variations of the $\eta$ Chamaleontis members.} 
\label{tab2}      
\centering         
\scriptsize 
\begin{tabular}{lccccccccc}     % 8 columns 
\hline\hline       
Source & RECX &  Ext. radius & Time bin  & Type & Factor & Flare dur. \\ 
              &              &  (arcsec)    &  (s)           &                 &            &  (ks) \\
\hline                    
EH Cha & 3   & 15 & 1500 &                     &  \\
EI Cha  &  4   & 15 &  900 & Flare           & 2.7 & $>$10\\
EK Cha & 5   & 15 & 1200 &                     & \\
EL Cha & 6   &  15 &  900 & Flare           & 2.6 & $\sim$4 \\
EM Cha & 7   &  15 & 900 & Variable     & \\
RS Cha & 8   & 10 & 1000 & Flare          & 1.8 & $\sim$20 \\
EN Cha & 9   &  15 &  900 & Flare           & $>$7.3 & $>$40 \\
EO Cha & 10 &  23 &  900 & Variable    & \\
EP Cha & 11 &  32 &  900 & Flare           & 1.4/1.7 & $\sim$18/16\\
EQ Cha & 12 & 32 & 1200 & Flare?    & \\
ES Cha & 14 & 8 & 1200 &                     & \\
ET Cha & 15 & 12 & 1800 & Flare?        & \\
\hline                  
\end{tabular}
\end{table}
%_____________________________________________________________

\subsubsection{Flares}
\label{flares}

During the exposure time, five stars showed clear flare activity (see Fig.~\ref{f3}).
Flares detected in our observation are not very intense in terms of 
count-rate enhancement (see Table~\ref{tab2}). Count-rate increase factors of 
approximately $2-3$ were detected, except for EN Cha (RECX 9). These values are 
similar to those observed in the less energetic flares detected in Taurus \citep{fra07} 
and in M field stars \citep{rob05,lop07,cre07}. In this sense, flares detected in 
members of $\eta$ Chamaleontis are similar to those observed in other 
M-type stars. 

We studied plasma parameters during the flares. In Table~\ref{tab3} we give
the results of the fits of hot plasma models to the X-ray spectrum of the stars during the 
flare and the \textit{quiescent} state for each star (assumed to be the characteristic level 
shown by the star previously to the flare, except for RECX 9, for which we used the last
6 ks of observation). 
As in the previous subsection, for the fit we used the XSPEC spectral fitting package 
with the APED and the WABS model (see Section~\ref{corona}). A $1T$--model was 
used for the quiescent state, except for RECX 4 and RECX 11, for which it was necessary 
to add a second thermal component to fit the spectrum. Note that they are the sources 
with the higher mean count-rates. The goodness of the fit is given in the table as the 
reduced $\chi^2$. For the flares, the fit was done after subtracting the quiescent spectrum.
An absorbed $1T$--model was then used, maintaining the same values of $N_\mathrm{H}$
and $Z$ than for the quiescent state. Note that temperatures obtained here are mean 
values, since we are integrating the entire event. 
The X-ray luminosities for the quiescent and the flare given in Table~\ref{tab3} are corrected 
for absorption. A mean distance of 97 pc was used to transform fluxes to luminosities for each
star. $L_\mathrm{X}^\mathrm{fl}$ and $\log E_\mathrm{X}^\mathrm{fl}$ are, respectively, the 
X-ray luminosity and liberated energy during the flare event. 
Fluxes, luminosities and energies were determined in the [0.3--8.0] energy band.
Note that our results for the flare parameters are relative to the pseudo-quiescent level 
chosen for the analysis. In particular, RECX 9 is observed during a flare decay phase 
and we used the last 8 ks of exposure as its quiescent level. 
In general, the values obtained here for the flare energies are consistent with those found 
for members of Taurus (see below).

\begin{figure}
   \centering
   \includegraphics[width=9cm]{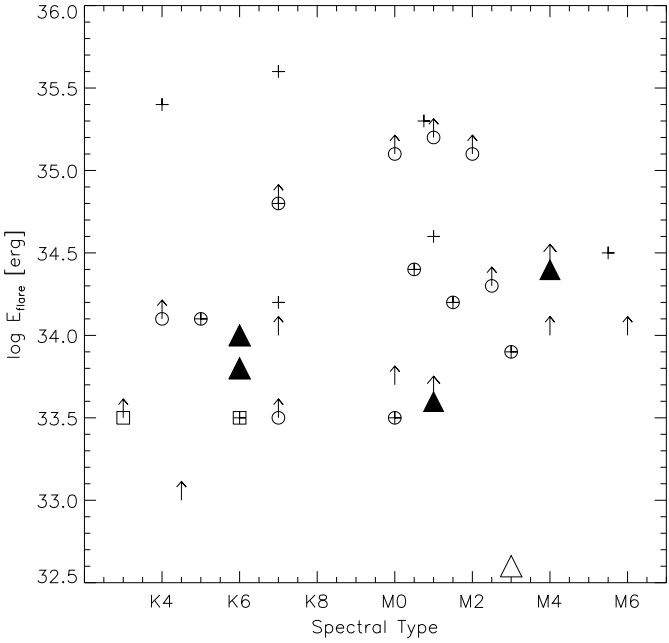}
      \caption{Energy liberated during the flare in the [0.3--8.0] energy band for 
      the stars in our sample (large triangles) and the stars of the XEST \citep[plusses;][]{ste07}.
      Upward arrows denote lower limits. Filled triangles are stars with 
      a disk in $\eta$ Chamaleontis. Stars classified as classical T Tauri in \citet{gud07}
      are marked with a large symbol (square or circle).}
       \label{f5}
   \end{figure}

Flare energies determined for our stars are typical of pre-main-sequence M stars. 
There seems to be none correlation with spectral-type, binarity or the presence of disks. 
The energy liberated during the flare in RECX 6 (the only star with 
no signs of disk among this group of stars showing flares)
is one order of magnitude lower than for the others. However, this result is not determining
because of the small sample we are using. We have compared our results with those 
obtained for the Taurus region by \citet{ste07}. In Fig.~\ref{f5} we plot flare 
energies observed for the stars in our sample (triangles) compared with those in 
the \textit{XMM-Newton} Extended Survey of the Taurus molecular cloud 
(XEST; plusses). Filled triangles and large symbols are used to mark stars with 
disks in our sample and the XEST one, respectively. Except for RECX 6 (with a 
flare energy $\log E_\mathrm{flare}$ [erg] = 32.6), the stars in our sample 
show flare energies similar to those in the XEST with the same spectral type. 
%There is a slight trend of increasing the flare energy with decreasing stellar mass 
%for M-type stars, but it seems to 
%be related to a bias in the observations due to the lower overall X-ray 
%emission in low-mass stars that would have prevented their detection
%(note that this behavior is observed only as a lower limit for flare energies of 
%Taurus sources, while we detected a lower energy for a flare in RECX 6).
No difference between stars with and without disk is observed. The stars marked 
with a circle in Fig.~\ref{f5} are classified as classical T Tauri in \citet{gud07}
and the stars marked with a square are embedded stars. 

%_____________________________________________________________
%                                             One column Table 
%_____________________________________________________________
%
\begin{table*}[!t]
\caption{Best fit values of the spectral model parameters for the quiescent 
and the flare. Errors are 90\% confidence level. X-ray luminosities and 
flare energies are unabsorbed values. We assumed a distance of 97 pc 
for all the stars. Fluxes, luminosities and energies were determined in the 
[0.3--8.0] energy band.} 
\label{tab3}     
\scriptsize
\centering          
\begin{tabular}{lccccccccccccc}     % 14 columns 
\hline\hline       
 & \multicolumn{7}{c}{Quiescent} & & \multicolumn{5}{c}{Flare} \\
\cline{2-8}\cline{10-14}
\noalign{\smallskip}
Source & $N_\mathrm{H}$ & $Z$ & $kT_1$ & $kT_2$ & $EM_1/EM_2$ & 
$\chi_\mathrm{q}^2$ & $\log L_\mathrm{X}^\mathrm{q}$ &  &
$kT_\mathrm{fl}$ & $EM_1/EM_\mathrm{fl}$ & $\chi_\mathrm{fl}^2$ & 
$\log L_\mathrm{X}^\mathrm{fl}$ & $\log E_\mathrm{X}^\mathrm{fl}$ \\
 RECX & ($\times 10^{21}$ cm$^{-2}$) & ($Z_\odot$) & (keV) & (keV) & &
  & (erg\,s$^{-1}$) & & (keV) & & & (erg\,s$^{-1}$) & (erg) \\
\hline                    
\noalign{\smallskip}
  4           & 0.3$_{-0.8}^{+1.4}$ & 0.13$_{-0.02}^{+0.03}$ & 0.29$_{-0.02}^{+0.02}$ & 
                  0.91$_{-0.04}^{+0.04}$ & 0.99 & 1.10 & 29.8 & & 2.27$_{-0.48}^{+0.44}$ &
                  0.77 & 1.07 & 29.9 & $>$33.6 \\
\noalign{\smallskip}
  6           & 0.0$_{-0.0}^{+0.2}$ & 0.08$_{-0.03}^{+0.02}$ & 0.73$_{-0.06}^{+0.04}$ & 
                  ...                                   & ...     & 0.84 & 29.3 & & 3.17$_{-1.83}^{+24.8}$ &
                  5.54 & 0.81 & 28.8 & 32.6 \\
\noalign{\smallskip}
  8           & 0.3$_{-0.1}^{+0.4}$ & 0.05$_{-0.02}^{+0.03}$ & 0.74$_{-0.11}^{+0.08}$ & 
                  ...                                   & ...     & 1.09 & 29.7 & & 1.05$_{-0.32}^{+0.33}$ &
                  3.94 & 0.75 & 29.2 & 33.6 \\
\noalign{\smallskip}
  9           & 1.6$_{-0.7}^{+0.4}$ & 0.13$_{-0.02}^{+0.13}$ & 1.24$_{-0.21}^{+0.37}$ & 
                  ...                                   & ...     & 0.70 & 29.2 & & 1.54$_{-0.37}^{+0.35}$ &
                  0.32 & 1.08 & 29.8 & $>$34.4 \\
\noalign{\smallskip}
11 (1$^\mathrm{st}$ flare) & 1.0$_{-0.3}^{+0.2}$ & 0.12$_{-0.03}^{+0.04}$ & 0.26$_{-0.01}^{+0.02}$ & 
                  1.21$_{-0.06}^{+0.06}$ & 1.70 & 1.10 & 30.5 & & 2.12$_{-0.58}^{+1.08}$ &
                  6.09 & 1.15 & 29.8 & 33.8 \\
\noalign{\smallskip}
11 (2$^\mathrm{nd}$ flare) & 1.0$_{-0.3}^{+0.2}$ & 0.12$_{-0.03}^{+0.04}$ & 0.26$_{-0.01}^{+0.02}$ & 
                  1.21$_{-0.06}^{+0.06}$ & 1.70 & 1.10 & 30.5 & & 1.49$_{-0.35}^{+0.55}$ &
                  3.76 & 1.11 & 30.0 & 34.0 \\
\noalign{\smallskip}
\hline                  
\end{tabular}
\end{table*}
%_____________________________________________________________

%_____________________________________________________________
% _____________________________________________________________

\subsubsection{X-ray emission modulation}

Six stars in our sample showed some kind of modulation in their light curves during 
the observations. These stars have RECX numbers 6, 7, 10, 11, 12, and 15. In RECX 6 
and RECX 11, the observed variations may be related to the decay of precedent flares. 
In particular, RECX 11 seems to show two flares during the observing period. In the 
remaining stars, the variations may be related to rotational modulation. 

The particular cases of RECX 7 and RECX 10 are interesting since they show a decrease
in count-rate during a short period of time ($\sim$3 and 1.5 hours respectively).  This
decrease could be related to an occultation of the corona by a companion. Of them, 
RECX 7 is known to be a double-line binary \citep{mam99} with a period of 2.6 days
\citep{lyo03}. However, RECX 10 has none known companion \citep{gue07}.

\section{Candidate cluster members in the field of view}
\label{candidates}

%_____________________________________________________________
   \begin{figure*}
   \centering
   \includegraphics[width=9cm]{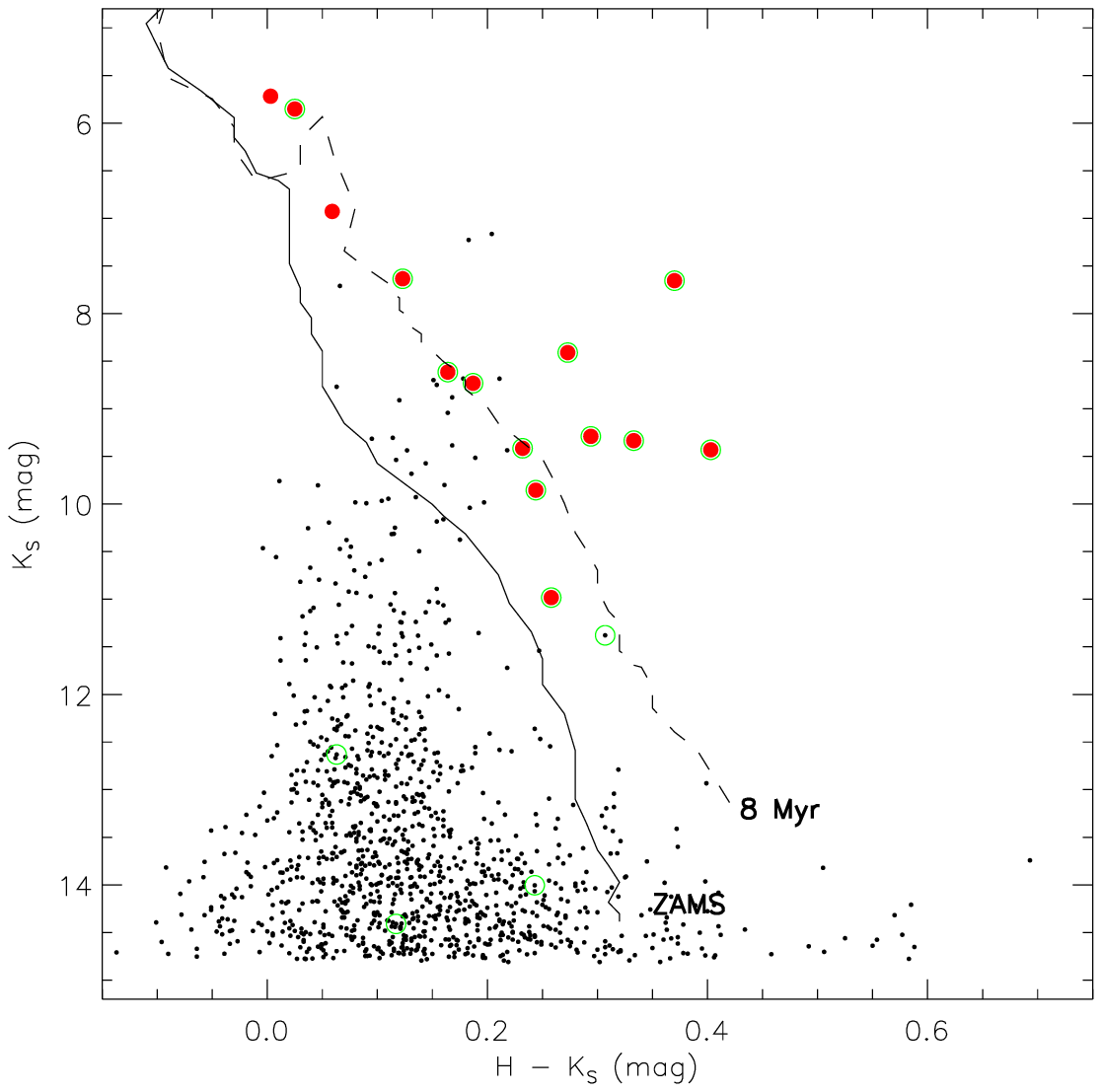}
   \includegraphics[width=9cm]{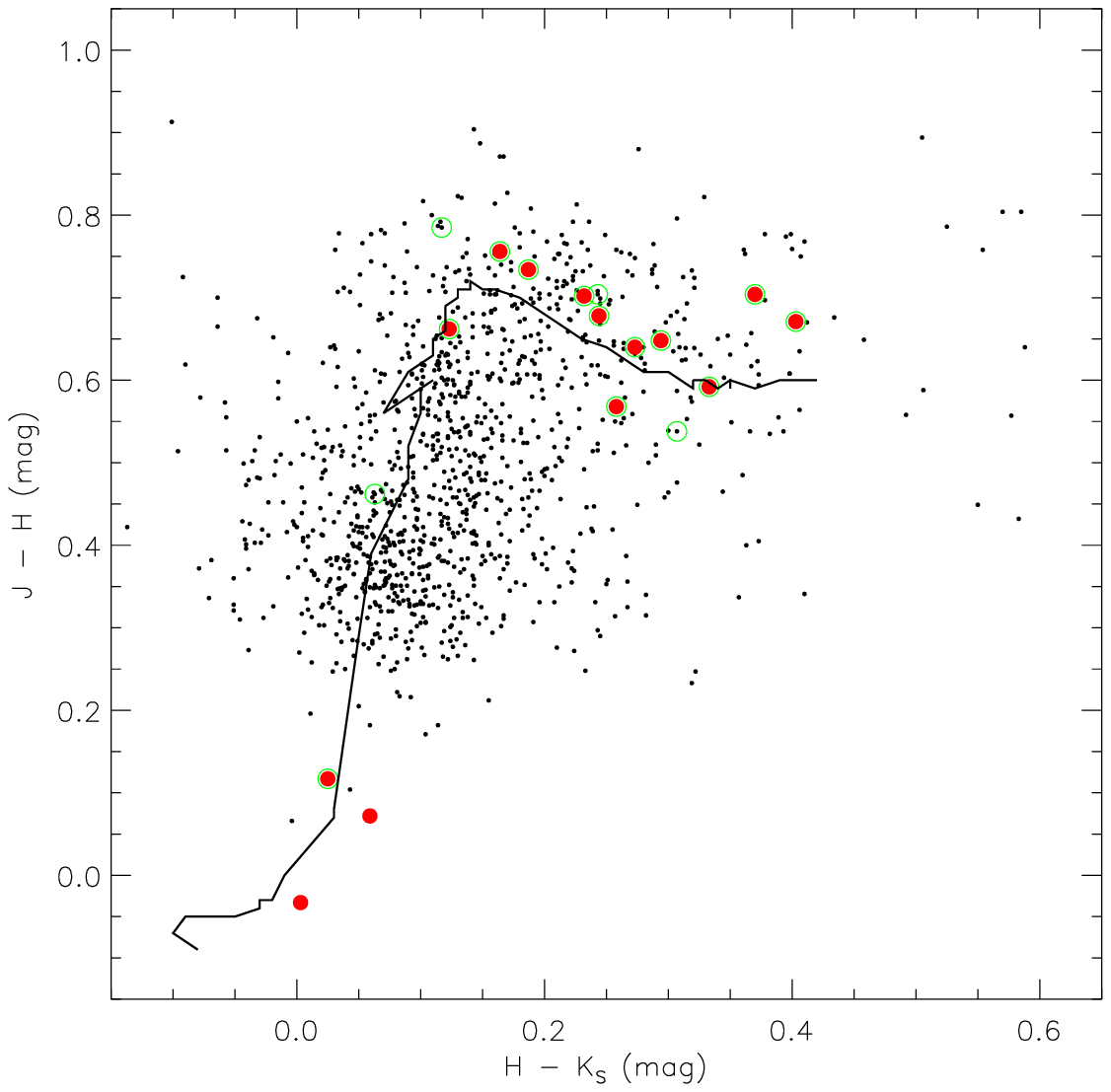}
      \caption{2MASS color-magnitude and color-color diagrams of the X-ray 
      sources detected in the XMM-Newton image, plus $\eta$ Cha. Filled circles 
      are the known cluster members in the EPIC field of view. Those members 
      detected in our observation are marked with a larger circle. Open circles 
      are X-ray sources in our list of detections. Small dots are 2MASS 
      sources in the field of view. ZAMS and 8 Myr isochrones are a combination 
      of pre-main sequence models by \citet{sie00}, for intermediate-mass stars 
      and \citet{bar98}, for low-mass stars. In the color-color diagram, we plot
      only the 8 Myr isochrone.}
       \label{f6}
   \end{figure*}
% _____________________________________________________________

Because of the low flux limit of our observation, we expected to detect all the cluster 
members emitting in X-rays in the EPIC field of view (see Section~\ref{obs}). 
Based on optical photometry, \citet{mam00} selected 50 candidates in the cluster 
core region. Later, \citet{law02} confirmed two low-mass stars inside the \textit{ROSAT} 
field of view to be members of $\eta$ Chamaleontis using optical spectroscopy. They 
were not detected during the \textit{ROSAT} observations. These two stars have 
RECX numbers 14 and 15 and have been both detected in our observation 
(see Section~\ref{obs} and Table~\ref{tabA1}). The searching radius for candidates 
has been increasing since then up to 5.5 deg \citep{luh04,lyo04,son04,lyo06,mur10}.
Six new members and 3 possible members have been identified at distances of 
between 1 and 6 deg from the cluster core. Most of these surveys were centered 
on finding very low mass stars and brown dwarfs cluster members
\citep[e.g.][]{luh04,lyo06,mur10}.

We searched for intermediate-mass (late-F and G type stars) members of the 
$\eta$ Chamaleontis cluster that, eventually, would had gone unnoticed in low-mass 
and brown dwarfs surveys. Those stars should emit in X-rays with fluxes above our 
detection limit ($f_\mathrm{X} \sim 1-3 \times 10^{-15}$ erg\,cm$^{-2}$\,s$^{-1}$; 
see Section~\ref{obs}). We first cross-correlated our X-ray sample with the 2MASS
database \citep{skr06}. 
A searching radius of 4 arcsec was used to prevent \textit{false} positives 
in the identification. \citet{cec04} estimated a positional error of the X-ray sources of 
the XMM-Newton Bright Serendipitous Survey of 6 arcsec. However, the mean distance
between known members cluster and its X-ray counterpart in our observation is 
$\sim 1.8$ arcsec with a maximum separation of 2.8 arcsec for RECX 11. A searching
radius of 4 arcmin is a conservative value. 

The results of cross-correlating our X-ray sample and the 2MASS
database are shown in Fig.~\ref{f6}. Small dots in the figure are 2MASS sources 
in the field of view of EPIC. Filled circles are the fourteen known cluster members in the 
same field of view. The twelve members detected in this observation are marked with a 
larger circle. Open circles are other X-ray sources in the field with a 2MASS counterpart. 
ZAMS and 8 Myr isochrone are also plotted. From these figures, only one 
of these other X-ray sources could be classified as possible member of the cluster. 
The star (2MASS J08483486-7853513) has near-infrared colors of an M4/6 dwarf, which 
would correspond to a mass of 0.10--0.13 M$_\odot$ for a cluster member. However, 
\citet{law02} already studied this star and found that it is likely a high-proper-motion 
dM5e foreground star at a distance of $\sim 50$ pc. The other three sources are %probably 
background stars (see Fig.~\ref{f6} and Table~\ref{tab4}). 
Note that there is only one star in the field that could fulfill the requirement to be an
intermediate-mass (FG-type star) cluster member in terms of near-infrared colors. This 
star is HD 76144. Nevertheless, the Hipparcos catalog gives a distance $d = 140$\,pc, 
quite larger than the value estimated for the cluster members. The star is neither 
detected in the \textit{XMM-Newton} observation with our detection procedure (see
Section~\ref{obs}). Therefore, we neglected also this star as a member of $\eta$
Chamaleontis.

%_____________________________________________________________
%                                             One column Table 
%_____________________________________________________________
%
\begin{table}[!t]
\caption{Other X-ray sources in the field with 2MASS counterpart non-members
of $\eta$ Chamaleontis.} 
\label{tab4}      
\scriptsize
\centering          
\begin{tabular}{ccccc}     % 5 columns 
\hline\hline       
2MASS & $J$    & $H$    &  $K_\mathrm{S}$ & Note \\ 
             & (mag) & (mag) & (mag) \\
\hline                    
J08394669-7900026 & 17.21 $\pm$ 0.22 & 16.42 $\pm$ 0.22 & 15.56 $\pm$ 0.23 & Galaxy \\
J08431595-7853422 & 14.95 $\pm$ 0.03 & 14.25 $\pm$ 0.04 & 14.00 $\pm$ 0.06 & M-type \\
J08440921-7906156 & 15.31 $\pm$ 0.05 & 14.53 $\pm$ 0.06 & 14.41 $\pm$ 0.07 & K-M giant \\
J08451844-7854426 & 13.16 $\pm$ 0.03 & 12.69 $\pm$ 0.02 & 12.63 $\pm$ 0.03 & K-type \\
J08483486-7853513 & 12.22 $\pm$ 0.02 & 11.69 $\pm$ 0.03 & 11.40 $\pm$ 0.02 & dM5e\tablefootmark{*} \\
\hline                  
\end{tabular}
\tablefoot{
\tablefoottext{*}{\citet{law02} rejected this star as a cluster member and gave 
spectral type M5 with emission. \citet{luh04} confirmed this result later.}
}

\end{table}
%_____________________________________________________________

\section{Notes on individual stars}
\label{notes}

\subsection{$\eta$ Cha (RECX 2)}

This star is the most massive cluster member (spectral type B8). It was detected by 
\citet{mam99} in a \textit{ROSAT}-HRI observation. From the observed X-ray flux, they 
inferred an X-ray 
luminosity $\log L_\mathrm{X} \mathrm{[erg\,cm^{-2}\,s^{-1}]} = 28.8$ in the HRI energy 
band. \citet{mam00} later suggested that $\eta$ Cha is likely a binary star, based on the 
observed variations in its radial velocity \citep{bus61}. Therefore, they argued that the 
X-ray emission is likely to be produced in a low-mass secondary, 
for which \citet{lyo04} estimates a mass of 0.5\,M$_\odot$ using 
$L_\mathrm{X}-M$ relationships.

$\eta$ Cha was not detected in our \textit{XMM-Newton} observation. At our detection 
flux limit $f_\mathrm{X} \sim 1-3 \times 10^{-15}$ erg\,cm$^{-2}$\,s$^{-1}$ in the energy 
band [0.3-8.0] keV --\,which is traduced in a luminosity 
$L_\mathrm{X} \sim 2 \times 10^{27}$ erg\,s$^{-1}$ at a distance of 100 pc\,-- any 
low-mass companion should have been detected (although see Section~\ref{variability}, 
where we suggest that ET Cha has a quiescent X-ray emission lower than the detection
limit).
The lack of X-ray emission from this star during the entire exposure is not simply explained 
in terms of variability.
A possible scenario would be one in which the primary (non-emitting) star would 
eclipse the X-ray emitter for long periods of time. Nevertheless, 
the fact that the X-ray emission would had gone unnoticed during this observation 
should not be discarded. Note that ES Cha (RECX 14), the member with the lowest 
mass in the field of view, was observed at the detection limit. A lower mass star 
(spectral type latter than M6) or even a brown dwarf may be part of this system.
Any of these scenarios should be checked through a robust determination of the 
system orbital parameters.

\subsection{RS Cha (RECX 8)}

RS Cha is a well-known Herbig Ae double-lined eclipsing binary %\citep[e.g.][]{boh09} 
with an orbital period $P_\mathrm{orb} = 1.67$ days, located in the instability strip
\citep{mar98, ale05}.
%\citep{ale05}. These authors showed that the system is located inside the instability strip 
%predicted by \citet{mar98}. Later, \citet{boh09} confirmed the pulsating nature of both stars 
%in the system.
The pulsating nature of both stars in the system was confirmed by \citet{boh09}.

The X-ray emission detected with \textit{ROSAT} by \citet{mam99} has been suggested
to come from a tertiary companion \citep{mam00,lyo04}. However, \citet{ale05}
found evidences that the time for the first conjunction of this system changes with 
time very rapidly, what is not expected for the presence of a third body in the system.
%found that the changes observed in the time for the first conjunction between 1975 
%and 2002 seem too large to be due to a third body in the system. 
Instead, the soft X-ray emission may come from the A-type stars themselves, especially 
if they have disturbed atmospheres \citep{gom09}. But it has been shown that
X-ray spectral properties of Herbig Ae/Be stars are similar to those observed
in late-type stellar companions to other Herbig Ae/Be stars \citep[e.g.][]{ste06,ste09}.
In fact, the X-ray luminosity of RS~Cha is similar to that observed for the early-type M 
members of the cluster (see Fig.~\ref{f2}).

\begin{figure}
   \centering
   \includegraphics[width=9cm]{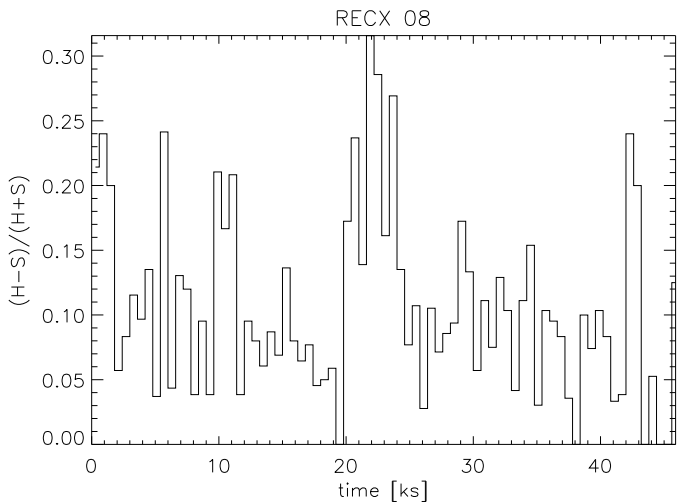}
      \caption{Evolution of the hardness-ratio of RS Cha with time. The soft ($S$) and 
      hard ($H$) energy bands are defined in the ranges [0.3-0.8] and [2.0-7.5] keV, 
      respectively. Each bin is 10 minutes long.}
       \label{f7}
   \end{figure}

In the \textit{XMM-Newton} observation, RS Cha underwent a flare. This fact is confirmed 
by the temporal evolution of the spectrum hardness-ratio (Fig.~\ref{f7}). Besides, from
the ephemeris for the primary minimum given by \citet{cla80} and the linear variation
of this relation with time observed by \citet{ale05}, we estimate the primary eclipse 
finished $\sim 2.3$ hours after the beginning of the exposure, more than three hours 
before the flare occurred. Therefore, the shape of the light curve cannot be produced by
an occultation of the X-ray emitter. 

The flare temperature determined from the X-ray spectrum is quite low and the amount 
of material involved in the flare event is not very large compared with the quiescent 
(see Table~\ref{tab3}). In this sense, the event detected in RS Cha is not very different
from those observed in late-type cluster members as EP Cha (RECX~11). 

%_____________________________________________________________
% _____________________________________________________________

\subsection{EM Cha and EO Cha (RECX 7 and RECX 10)}

During the observation, both stars presented a decrease of their count-rate by a factor 
of $\sim 1.5$ during 3 and 1.5 hours, respectively, in a way that resembles the occultation 
of part of the stellar corona by a companion (see Fig.~\ref{f3}). EM Cha is known to be a 
binary star with a mass ratio of approximately 2.3:1 \citep{lyo04}. Following the results
of Section~\ref{corona}, the low-mass companion may have an X-ray luminosity 
of 0.5--1 orders of magnitude lower than the primary (see Fig.~\ref{f2}). The occultation 
of part of the corona of the primary star by the secondary would produce a decrease 
of the observed total flux (i.e. of the observed count-rate). EO Cha, however, shows 
none sign of binarity \citep{gue07,lyo04}. Therefore, the eclipse scenario does not 
seem probable for EO Cha. 

The observed light curves may also be explained, in both cases, 
by the occultation of an active region behind the visible stellar hemisphere due to 
stellar rotation. In fact, the count-rate of EO Cha increased during approximately 
2.5 hours before the rapid decrease during 1.5 hours at the middle of the exposure. 
This fact may be interpreted as the occultation of an active region by the star's
disk. 
Nevertheless, with the available data, none of these two scenarios can be 
discarded. 

\subsection{EN Cha (RECX 9)}

EN Cha is a classical T Tauri, binary system \citep{koh02,sic09} formed by two nearly 
identical M-type stars \citep{mam00,lyo04}. During the \textit{ROSAT} observation,
this star was marginally detected above the background level, with an X-ray luminosity 
$\log L_\mathrm{X} \mathrm{[erg\,cm^{-2}\,s^{-1}]} = 28.5$ \citep{mam00}. 
During the \textit{XMM-Newton} observation, the star showed a luminosity 
one order of magnitude higher than during the \textit{ROSAT} observation.

From the light curve and coronal parameters obtained from the plasma model 
fitting (Fig.~\ref{f3} and Table~\ref{tab2}), it seems clear that EN Cha underwent a 
flare during the \textit{XMM-Newton} observation. %According to the results presented
%in Fig.~\ref{f2}, the system quiescent emission level may be similar to the value
%observed by \citet{mam99,mam00}. 

\section{Summary and conclusions}
\label{summary}

In this paper we presented a detailed study of the X-ray emission properties
of the $\eta$ Chamaleontis cluster members, based on a deep \textit{XMM-Newton}
observation toward the cluster core. This study is complete in terms of cluster members. 
We detected all the (X-ray emitter) members of $\eta$ Chamaleontis down to the sub-stellar 
mass limit.
We determined X-ray luminosities, coronal temperatures, abundances
and column densities from hot plasma model-fitting. We also studied the 
variability of the sources. 

The comparison between the X-ray luminosities derived in this work and those obtained
by \citet{mam00} have shown that their values are overestimated by a factor of 2, on average. 
The coronal properties determined for cluster members are typical of highly active 
stars. Multi-temperature models with at least two components were needed in the
model-fitting. In the cases in which flare-like events were detected, a third hotter 
component was necessary to account for the enhancement of hard X-ray emission. 
In general, the stars that underwent a flare during our observations showed enhanced 
pseudo-quiescent X-ray luminosities with respect to members with similar spectral type
that showed no flares. 
Six flares were detected in five of the cluster members, with energies that are typical of 
pre-main- and main-sequence M stars. For these stars, coronal properties and X-ray 
luminosities were determined for the flare and the pseudo-quiescent states. 

From the comparison of the X-ray luminosity of cluster members in different environments:
the presence or not of a protoplanetary disk in different evolutionary stages and binarity, 
we concluded that the only parameter that seems to influence the overall X-ray emission 
of the stars in $\eta$ Chamaleontis is the spectral type. A similar conclusion may be 
extracted for the flare energies, although the sample of flare-like events during our 
observations is too poor to achieve a robust conclusion. 

To complete our study, we searched for other cluster members in the field-of-view that 
could had gone unnoticed in previous surveys. We found five candidates, but discarded
them as cluster members after a detailed study of their optical spectroscopic properties.

\begin{acknowledgements}
      J.L-S. and M.A.L-G. acknowledge support by the Spanish Ministerio de Ciencia 
      e Innovaci\'on under grant AYA2008-06423-C03-03. J.F.A.C., is researcher of 
      the CONICET and acknowledges support by grant PICT 2007-02177 (SecyT).
      We would like to acknowledge the anonymous referee for his/her useful 
      comments on the text content.
\end{acknowledgements}

\appendix

\section{On-line material}

\begin{table*}
\caption{Sources detected in the 0.3--7.5 keV energy band
using PWDetect with detection threshold $SNR = 5$.}
\label{tabA1}
\scriptsize
\label{spectra}\centering
\begin{tabular}{lcccccl}
\hline
Src \# & Right Ascension & Declination & Significance &
Count-rate & Observed flux$^{*}$ & Notes \\
        &     (h m s)     &   ($^\circ~\arcmin~\arcsec$) & &
($\times 10^{-3}$ s$^{-1}$) & ($\times 10^{-13}$ erg cm$^{-2}$ s$^{-1}$) \\
\hline
 1 &  08 45 33.9 & -79 13 33.5 &   6.0 &    1.3 $\pm$  0.7 &   0.09 $\pm$ 0.05 &  \\
 2 &  08 42 42.8 & -79 10 59.3 &   5.8 &    1.9 $\pm$  0.3 &   0.14 $\pm$ 0.03 &  \\
 3 &  08 44 07.3 & -79 10 13.1 &   6.1 &    1.1 $\pm$  0.2 &   0.08 $\pm$ 0.02 &  \\
 4 &  08 43 19.7 & -79 09 50.1 &   6.6 &    0.9 $\pm$  0.2 &   0.07 $\pm$ 0.02 &  \\
 5 &  08 40 53.4 & -79 09 03.9 &   6.5 &    3.8 $\pm$  0.4 &   0.29 $\pm$ 0.05 &  \\
 6 &  08 45 42.3 & -79 07 11.7 &   7.0 &    0.6 $\pm$  0.1 &   0.04 $\pm$ 0.01 &  \\
 7 &  08 44  9.4 & -79 06 17.8 &   5.9 &    0.6 $\pm$  0.1 &   0.04 $\pm$ 0.01 & 2MASS J08440921-7906156, field giant \\
 8 &  08 42 10.8 & -79 06 02.3 &   5.8 &    0.3 $\pm$  0.1 &   0.02 $\pm$ 0.01 &  \\
 9 &  08 43 18.4 & -79 05 20.2 &  14.0 &    1.4 $\pm$  0.2 &   0.10 $\pm$ 0.02 & RECX 15 \\
10 &  08 41 30.5 & -79 05 02.9 &   8.1 &    0.7 $\pm$  0.2 &   0.05 $\pm$ 0.01 &  \\
11 &  08 49 20.9 & -79 04 50.9 &   5.6 &    1.0 $\pm$  0.6 &   0.08 $\pm$ 0.05 &  \\
12 &  08 43 07.7 & -79 04 53.9 & 244.9 &  131.9 $\pm$  1.6 &   9.86 $\pm$ 1.32 & RECX 7 \\
13 &  08 42 49.6 & -79 04 31.4 &   5.3 &    0.5 $\pm$  0.1 &   0.03 $\pm$ 0.01 &  \\
14 &  08 43 12.5 & -79 04 13.3 & 140.1 &   45.7 $\pm$  1.0 &   3.41 $\pm$ 0.46 & RECX 8 \\
15 &  08 43 18.0 & -79 04 10.6 &   5.7 &    0.3 $\pm$  0.2 &   0.02 $\pm$ 0.02 &  \\
16 &  08 42 24.2 & -79 04 04.6 & 170.4 &   67.3 $\pm$  1.2 &   5.03 $\pm$ 0.68 & RECX 4 \\
17 &  08 47 15.0 & -79 03 55.0 &   6.1 &    1.5 $\pm$  0.2 &   0.12 $\pm$ 0.02 &  \\
18 &  08 41 37.2 & -79 03 31.3 &  43.2 &   10.3 $\pm$  0.5 &   0.77 $\pm$ 0.11 & RECX 3 \\
19 &  08 45 50.4 & -79 02 58.0 &   7.2 &    0.5 $\pm$  0.1 &   0.04 $\pm$ 0.01 &  \\
20 &  08 45 47.8 & -79 02 38.4 &   6.1 &    0.2 $\pm$  0.1 &   0.02 $\pm$ 0.01 &  \\
21 &  08 45 13.8 & -79 02 37.2 &   9.9 &    0.8 $\pm$  0.1 &   0.06 $\pm$ 0.01 &  \\
22 &  08 41 22.7 & -79 02 33.0 &   6.9 &    0.9 $\pm$  0.2 &   0.07 $\pm$ 0.02 &  \\
23 &  08 43 31.6 & -79 02 08.3 &  14.0 &    1.0 $\pm$  0.1 &   0.08 $\pm$ 0.01 &  \\
24 &  08 48 10.6 & -79 01 05.9 &   8.7 &    1.5 $\pm$  0.2 &   0.11 $\pm$ 0.02 &  \\
25 &  08 41 39.9 & -79 00 55.7 &   5.5 &    0.2 $\pm$  0.1 &   0.01 $\pm$ 0.01 &  \\
26 &  08 40 32.8 & -79 00 45.8 &   5.7 &    0.7 $\pm$  0.1 &   0.05 $\pm$ 0.01 &  \\
27 &  08 46 55.5 & -79 00 17.7 &   5.1 &    0.3 $\pm$  0.1 &   0.02 $\pm$ 0.01 &  \\
28 &  08 40 24.1 & -79 00 12.0 &   5.7 &    0.6 $\pm$  0.1 &   0.05 $\pm$ 0.01 &  \\
29 &  08 47 01.8 & -79 00 11.4 &   5.8 &    0.4 $\pm$  0.1 &   0.03 $\pm$ 0.01 &  \\
30 &  08 39 47.2 & -79 00 04.6 &  11.2 &    2.0 $\pm$  0.2 &   0.15 $\pm$ 0.03 & 2MASS J08394669-7900026, AGN \\
31 &  08 45 05.3 & -79 00 09.3 &  16.9 &    1.3 $\pm$  0.1 &   0.10 $\pm$ 0.02 &  \\
32 &  08 46 02.6 & -79 00 01.0 &   7.7 &    0.6 $\pm$  0.1 &   0.05 $\pm$ 0.01 &  \\
33 &  08 46 20.8 & -78 59 58.2 &   6.0 &    0.5 $\pm$  0.2 &   0.04 $\pm$ 0.02 &  \\
34 &  08 47 02.4 & -78 59 36.5 & 320.1 &  251.6 $\pm$  2.0 &  18.81 $\pm$ 2.51 & RECX 11 \\
35 &  08 41 36.3 & -78 59 34.3 &   6.5 &    0.6 $\pm$  0.1 &   0.04 $\pm$ 0.01 &  \\
36 &  08 43 20.7 & -78 59 34.9 &  13.7 &    0.7 $\pm$  0.1 &   0.05 $\pm$ 0.01 &  \\
37 &  08 44 17.1 & -78 59 10.0 & 149.4 &   31.1 $\pm$  0.5 &   2.33 $\pm$ 0.31 & RECX 9 \\
38 &  08 46 56.9 & -78 58 55.5 &   5.8 &    0.4 $\pm$  0.1 &   0.03 $\pm$ 0.01 &  \\
39 &  08 40 29.4 & -78 58 41.0 &  15.4 &    2.9 $\pm$  0.3 &   0.21 $\pm$ 0.04 &  \\
40 &  08 41 38.3 & -78 58 41.3 &  32.2 &    5.3 $\pm$  0.3 &   0.40 $\pm$ 0.06 &  \\
41 &  08 44 54.4 & -78 58 38.6 &  12.1 &    0.6 $\pm$  0.1 &   0.05 $\pm$ 0.01 &  \\
42 &  08 40 34.5 & -78 58 31.4 &  10.6 &    1.0 $\pm$  0.2 &   0.08 $\pm$ 0.02 &  \\
43 &  08 43 36.7 & -78 58 29.4 &   6.9 &    0.3 $\pm$  0.1 &   0.02 $\pm$ 0.01 &  \\
44 &  08 44 25.5 & -78 58 18.9 &  14.5 &    1.1 $\pm$  0.1 &   0.08 $\pm$ 0.01 &  \\
45 &  08 44 06.7 & -78 57 53.4 &  21.2 &    3.1 $\pm$  0.2 &   0.23 $\pm$ 0.03 &  \\
46 &  08 42 27.6 & -78 57 49.5 &  58.7 &    8.3 $\pm$  0.3 &   0.62 $\pm$ 0.09 & RECX 5 \\
47 &  08 43 24.5 & -78 56 46.1 &  15.8 &    1.0 $\pm$  0.1 &   0.07 $\pm$ 0.01 &  \\
48 &  08 43 36.1 & -78 56 32.7 &   5.4 &    0.1 $\pm$  0.0 &   0.01 $\pm$ 0.00 &  \\
49 &  08 41 09.0 & -78 56 01.0 &   7.8 &    0.8 $\pm$  0.1 &   0.06 $\pm$ 0.01 &  \\
50 &  08 49 01.1 & -78 55 52.3 &  22.3 &    8.2 $\pm$  1.3 &   0.62 $\pm$ 0.13 &  \\
51 &  08 44 42.3 & -78 55 51.2 &   7.9 &    0.6 $\pm$  0.1 &   0.04 $\pm$ 0.01 &  \\
52 &  08 43 39.0 & -78 55 41.4 &  23.6 &    2.1 $\pm$  0.2 &   0.16 $\pm$ 0.02 &  \\
53 &  08 45 36.0 & -78 55 35.8 &   6.2 &    0.6 $\pm$  0.1 &   0.05 $\pm$ 0.01 &  \\
54 &  08 48 01.1 & -78 55 20.0 &   7.8 &    0.9 $\pm$  0.2 &   0.07 $\pm$ 0.02 &  \\
55 &  08 47 57.4 & -78 54 52.9 & 128.4 &  167.2 $\pm$  1.7 &  12.50 $\pm$ 1.67 & RECX 12 \\
56 &  08 48 35.9 & -78 54 50.0 &  10.7 &    2.3 $\pm$  0.3 &   0.17 $\pm$ 0.03 &  \\
57 &  08 41 20.2 & -78 54 49.4 &   7.9 &    1.6 $\pm$  0.1 &   0.12 $\pm$ 0.02 &  \\
58 &  08 40 20.3 & -78 54 42.3 &  11.1 &    3.6 $\pm$  0.3 &   0.27 $\pm$ 0.04 &  \\
59 &  08 42 39.3 & -78 54 43.7 & 116.1 &   27.0 $\pm$  0.6 &   2.02 $\pm$ 0.27 & RECX 6 \\
60 &  08 45 19.6 & -78 54 43.8 &  10.7 &    0.9 $\pm$  0.2 &   0.07 $\pm$ 0.02 &  \\
61 &  08 45 01.9 & -78 54 40.4 &  11.9 &    1.0 $\pm$  0.1 &   0.08 $\pm$ 0.01 &  \\
62 &  08 47 52.2 & -78 54 16.4 &   6.2 &    0.5 $\pm$  0.1 &   0.03 $\pm$ 0.01 &  \\
63 &  08 43 13.7 & -78 54 14.5 &   6.5 &    0.5 $\pm$  0.1 &   0.04 $\pm$ 0.01 &  \\
64 &  08 44 29.9 & -78 54 03.5 &   5.3 &    0.2 $\pm$  0.0 &   0.01 $\pm$ 0.00 &  \\
65 &  08 48 35.4 & -78 53 48.4 &   5.9 &    1.6 $\pm$  0.4 &   0.12 $\pm$ 0.04 & 2MASS J08483486-7853513, field dM5e \\
66 &  08 43 16.6 & -78 53 44.1 &   7.6 &    0.5 $\pm$  0.1 &   0.03 $\pm$ 0.01 & 2MASS J08431595-7853422, field M \\
67 &  08 43 33.4 & -78 53 24.0 &  24.2 &    2.6 $\pm$  0.2 &   0.19 $\pm$ 0.03 &  \\
68 &  08 39 47.4 & -78 53 01.4 &  36.2 &   22.5 $\pm$ 11.3 &   1.68 $\pm$ 0.88 &  \\
69 &  08 41 29.9 & -78 53 05.2 &   5.1 &    0.6 $\pm$  0.1 &   0.04 $\pm$ 0.01 & RECX 14 \\
70 &  08 45 08.9 & -78 53 03.5 &   5.7 &    0.2 $\pm$  0.1 &   0.02 $\pm$ 0.00 &  \\
71 &  08 45 39.8 & -78 52 58.3 &   8.4 &    1.1 $\pm$  0.2 &   0.08 $\pm$ 0.02 &  \\
72 &  08 41 43.0 & -78 52 42.9 &  12.5 &    1.4 $\pm$  0.2 &   0.11 $\pm$ 0.02 &  \\
73 &  08 41 26.1 & -78 52 16.5 &  11.5 &    1.2 $\pm$  0.2 &   0.09 $\pm$ 0.02 &  \\
74 &  08 47 39.1 & -78 50 36.0 &   6.8 &    1.2 $\pm$  0.2 &   0.09 $\pm$ 0.02 &  \\
75 &  08 47 52.1 & -78 50 10.4 &   8.2 &    0.8 $\pm$  0.2 &   0.06 $\pm$ 0.01 &  \\
76 &  08 45 18.5 & -78 49 40.9 &   5.7 &    0.9 $\pm$  0.2 &   0.07 $\pm$ 0.02 &  \\
77 &  08 44 54.2 & -78 48 36.9 &   9.9 &    1.1 $\pm$  0.2 &   0.08 $\pm$ 0.02 &  \\
78 &  08 45 04.5 & -78 48 28.3 &  14.4 &    3.1 $\pm$  0.3 &   0.23 $\pm$ 0.04 &  \\
79 &  08 43 37.7 & -78 48 25.5 &   8.0 &    0.8 $\pm$  0.2 &   0.06 $\pm$ 0.01 &  \\
80 &  08 44 59.8 & -78 48 15.8 &  11.7 &    2.1 $\pm$  0.3 &   0.16 $\pm$ 0.03 &  \\
81 &  08 43 32.2 & -78 48 13.4 &  10.6 &    0.9 $\pm$  0.2 &   0.07 $\pm$ 0.01 &  \\
82 &  08 45 53.2 & -78 47 59.2 &   7.4 &    2.3 $\pm$  0.3 &   0.17 $\pm$ 0.03 &  \\
83 &  08 46 14.0 & -78 47 39.0 &   5.4 &    1.0 $\pm$  0.2 &   0.07 $\pm$ 0.02 &  \\
84 &  08 45 23.7 & -78 47 23.8 &  11.3 &    2.6 $\pm$  0.3 &   0.19 $\pm$ 0.03 &  \\
85 &  08 44 32.2 & -78 46 31.7 &  88.5 &   41.3 $\pm$  1.4 &   3.09 $\pm$ 0.42 & RECX 10 \\
86 &  08 45 27.8 & -78 45 45.4 &   5.7 &    1.0 $\pm$  0.2 &   0.07 $\pm$ 0.02 &  \\
\hline
\end{tabular}
\tablefoot{
\tablefoottext{*}{Observed fluxes determined using the conversion factor 
$CF = 1.5 \pm 0.2 \times 10^{-9}$ erg\,ph$^{-1}$. Fluxes from the 
spectral fitting for the known cluster members are given in Table~\ref{tab1}.}
}
\end{table*}

%_____________________________________________________________
\begin{figure*}[!t]
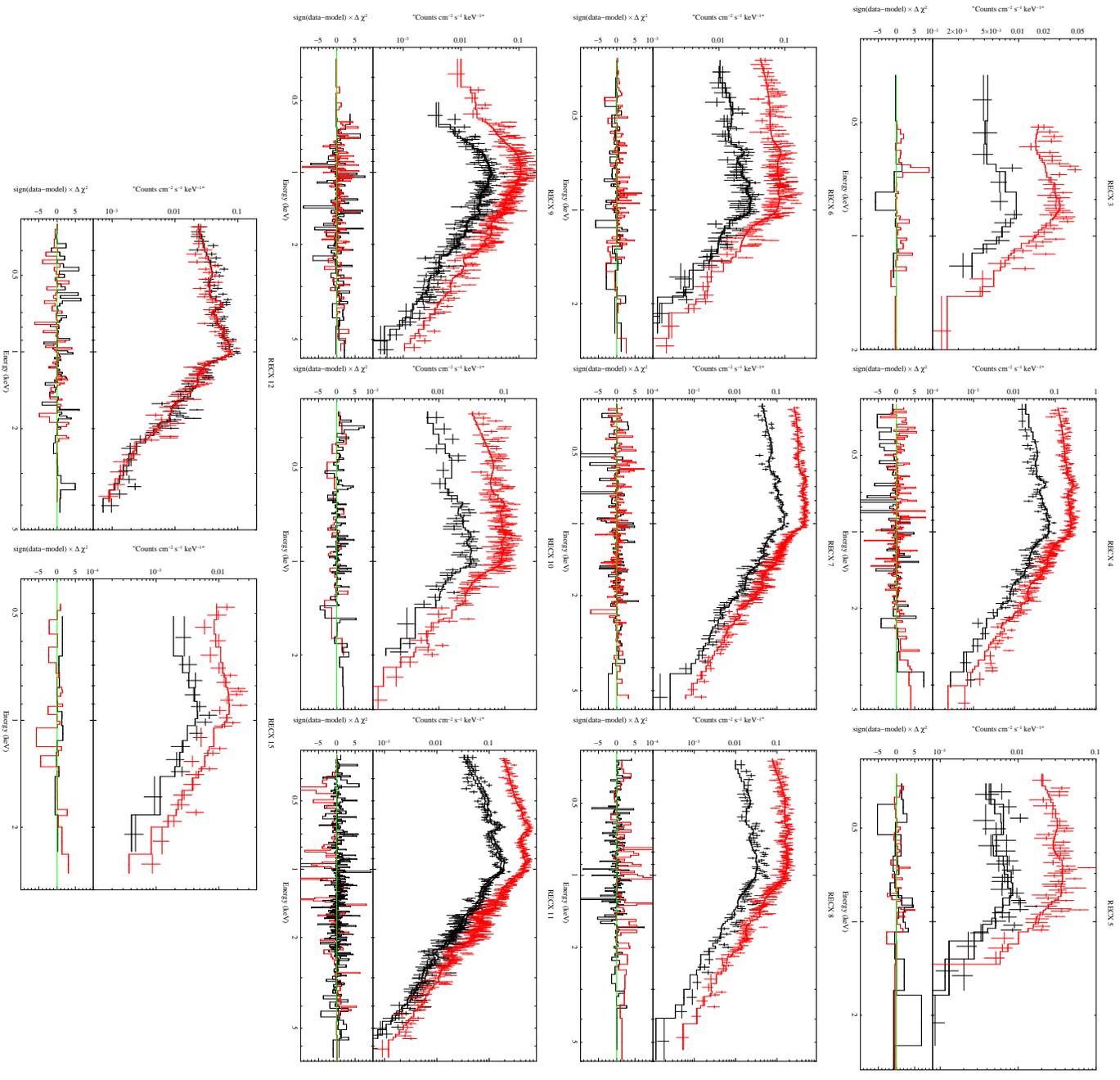

\centering
\includegraphics[width=4.5cm,angle=270]{acepted-recx3.ps}
\includegraphics[width=4.5cm,angle=270]{acepted-recx4.ps}
\includegraphics[width=4.5cm,angle=270]{acepted-recx5.ps}
\includegraphics[width=4.5cm,angle=270]{acepted-recx6.ps}
\includegraphics[width=4.5cm,angle=270]{acepted-recx7.ps}
\includegraphics[width=4.5cm,angle=270]{acepted-recx8.ps}
\includegraphics[width=4.5cm,angle=270]{acepted-recx9.ps}
\includegraphics[width=4.5cm,angle=270]{acepted-rec10.ps}
\includegraphics[width=4.5cm,angle=270]{acepted-recx11.ps}
\includegraphics[width=4.5cm,angle=270]{acepted-rec12.ps}
\includegraphics[width=4.5cm,angle=270]{acepted-rec15.ps}
   \caption{X-ray spectra for known members of $\eta$ Chamaleontis. Red is for 
   EPIC-pn and black is for EPIC-mos, except for RECX 12 that is located in an 
   EPIC-pn gap. The best fit to each spectrum is also plotted. The bottom panel 
   represents the deviation of the model from the observed spectra in each 
   spectral bin.}
   \label{fA1}
\end{figure*}
%_____________________________________________________________

%_____________________________________________________________
\begin{figure*}
\centering
\includegraphics[width=5.5cm]{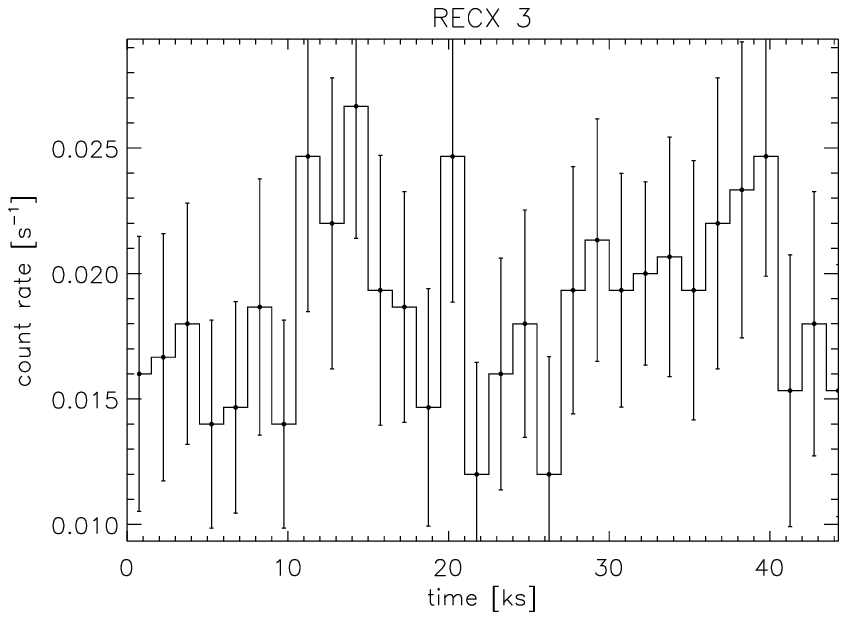}
\includegraphics[width=5.5cm]{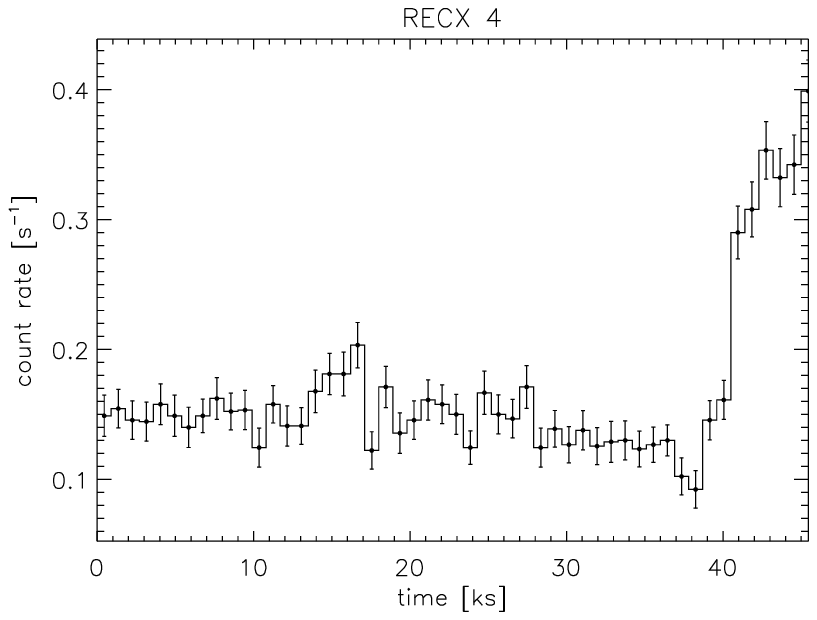}
\includegraphics[width=5.5cm]{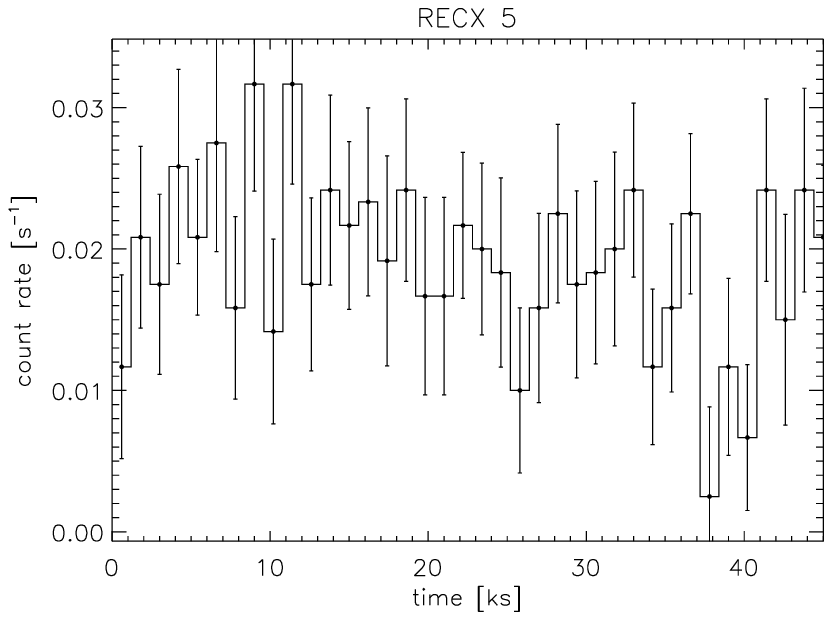}
\includegraphics[width=5.5cm]{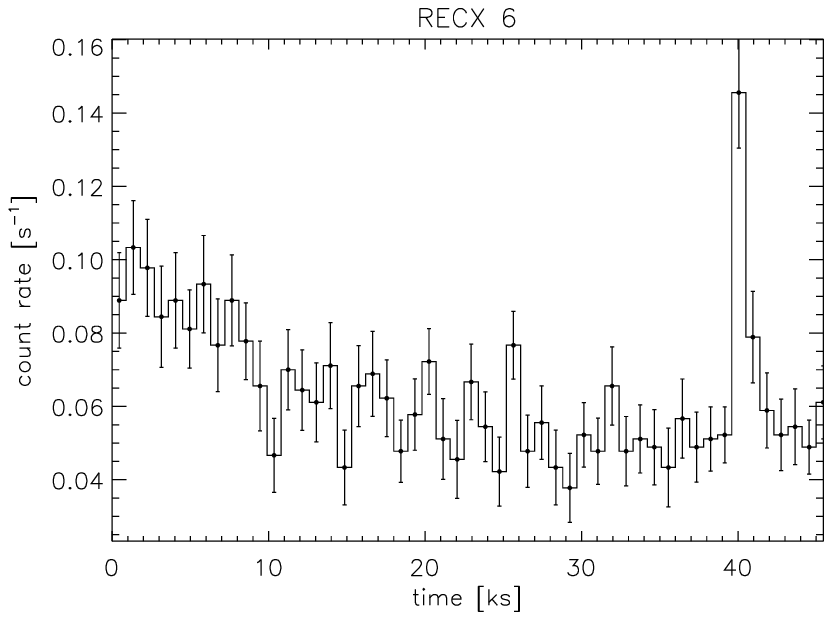}
\includegraphics[width=5.5cm]{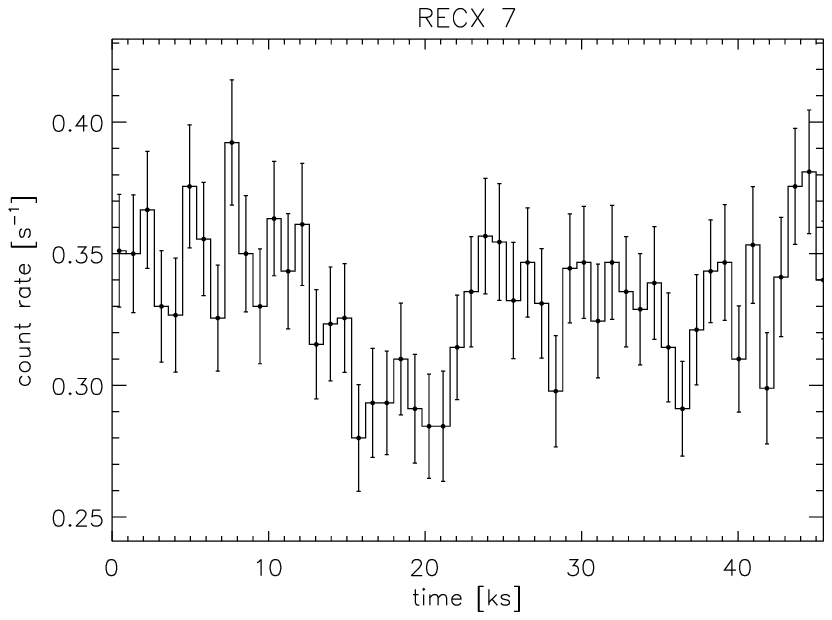}
\includegraphics[width=5.5cm]{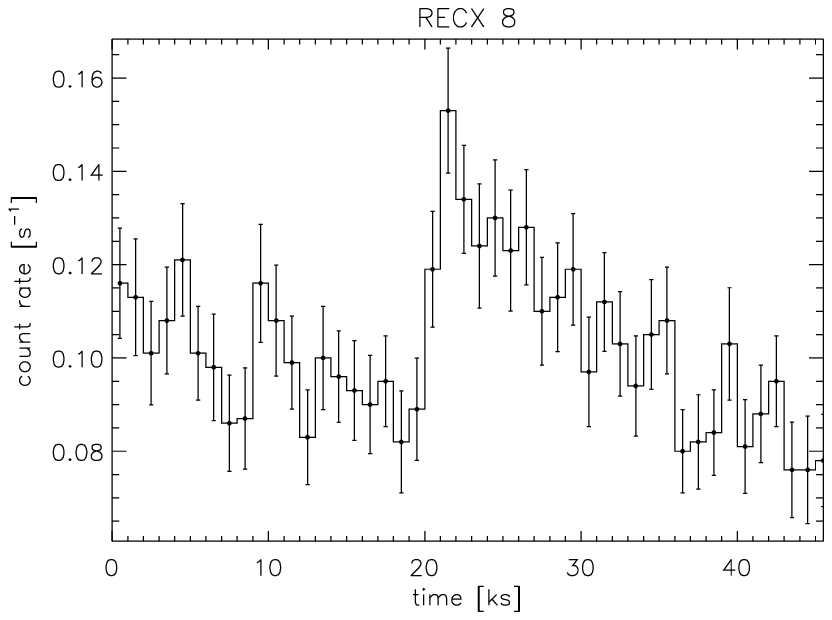}
\includegraphics[width=5.5cm]{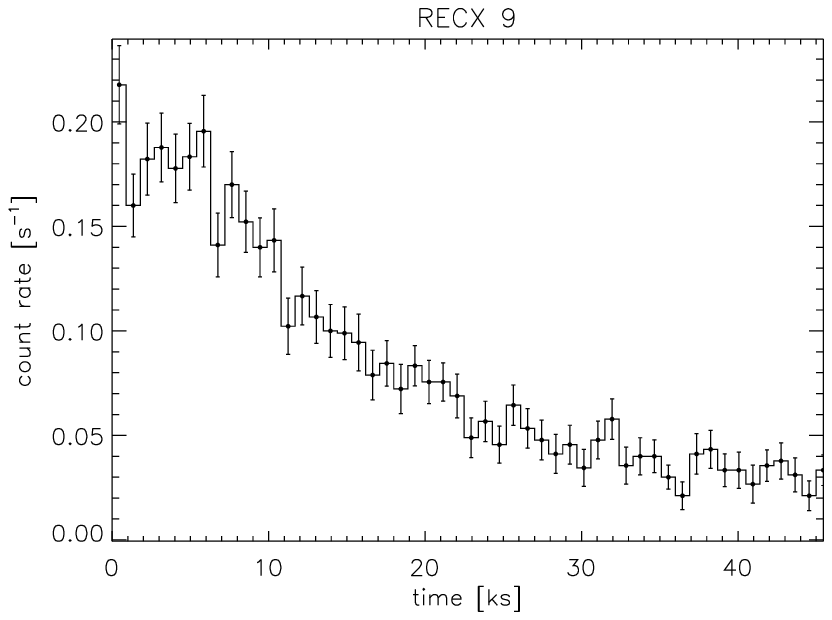}
\includegraphics[width=5.5cm]{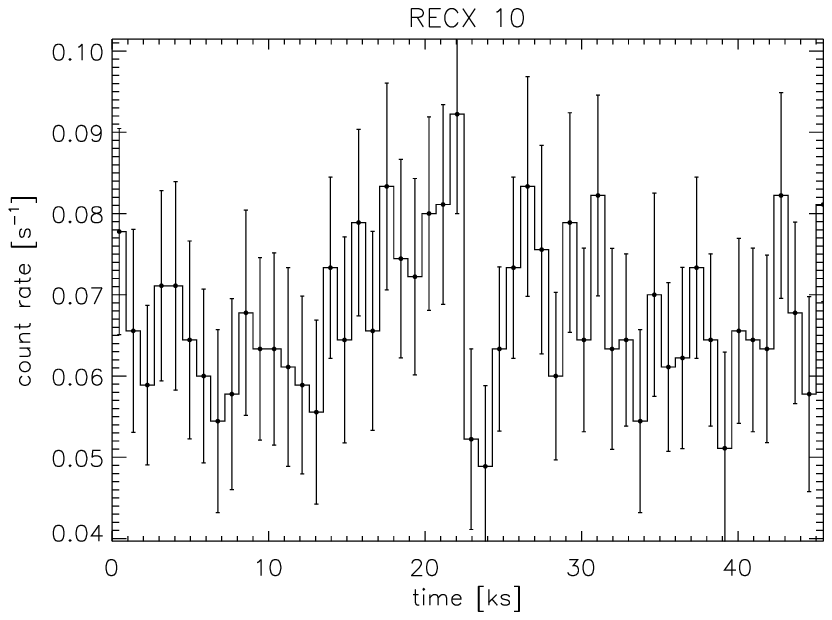}
\includegraphics[width=5.5cm]{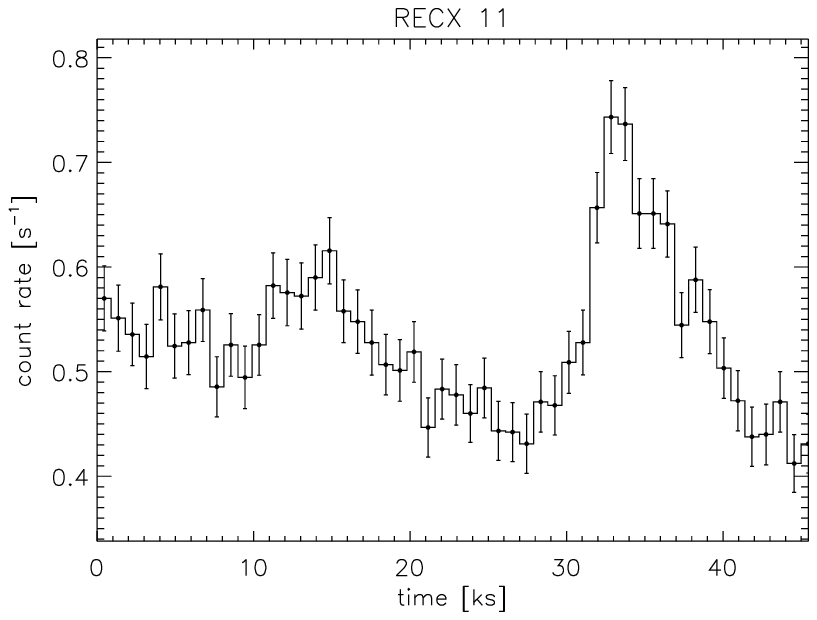}
\includegraphics[width=5.5cm]{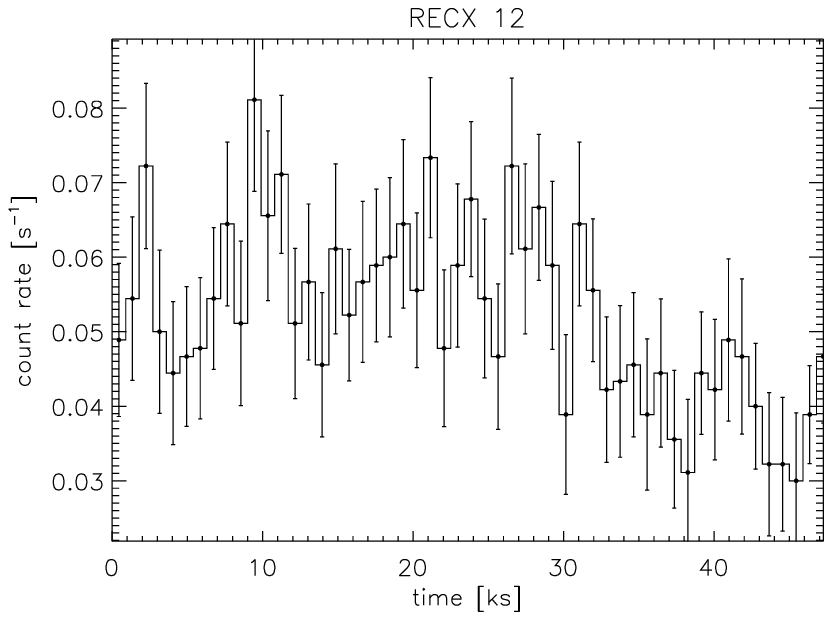}
\includegraphics[width=5.5cm]{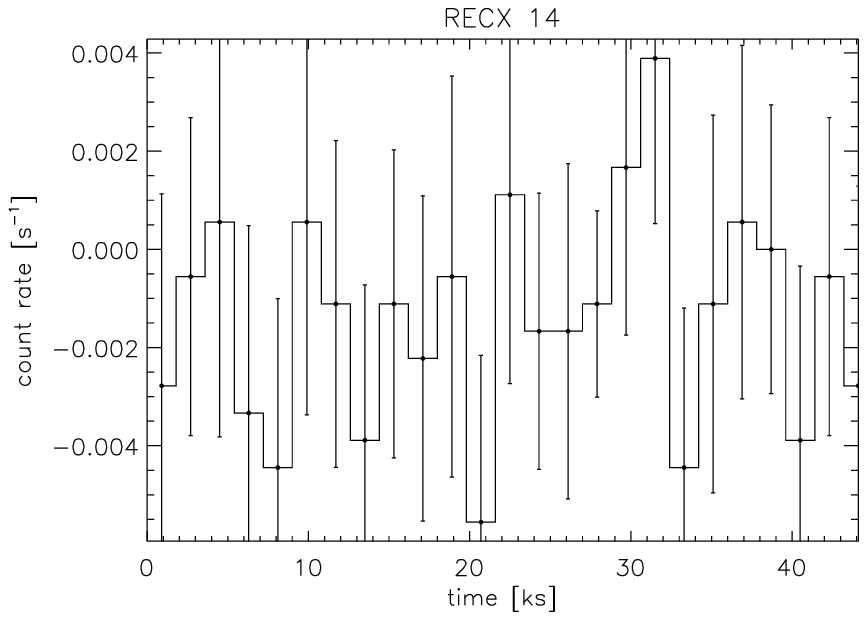}
\includegraphics[width=5.5cm]{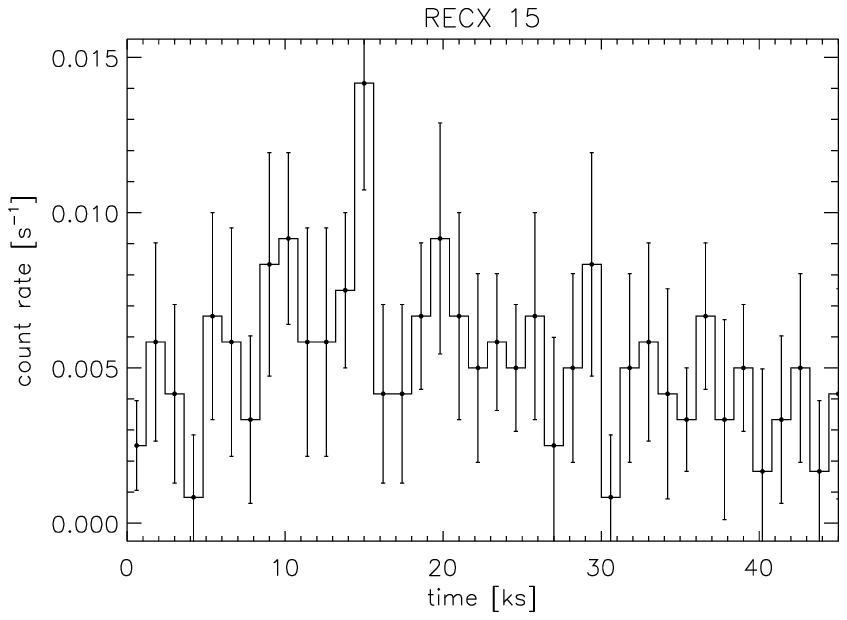}
   \caption{X-ray light curves in the energy band 0.3--8.0 keV for known members 
   of $\eta$ Chamaleontis. Time binning ranges from 15 to 30 minutes. 
              }
   \label{f3}
\end{figure*}
%_____________________________________________________________

\end{document}